\documentclass[twocolumn]{autart}    % Enable this line and disable the 
\usepackage{cite}	     	% for citations
\usepackage{newtxtext}  	% goes with newtxmath, for text, must include before 
\usepackage{textcomp}   	% compatibility
\usepackage[euler]{textgreek}
\usepackage{lipsum}	     	% for automatic text blocks
\usepackage[normalem]{ulem}

\usepackage{soul, color}   	% to highlight
\usepackage[svgnames]{xcolor} % to change text color
\usepackage[utf8]{inputenc} % to use foreign accents
\usepackage[T1]{fontenc}    % compatibility
\usepackage[shortlabels]{enumitem}  		% bettere enumerate
\usepackage{url}
\usepackage{amsmath}    	% allows further math commands
\usepackage{amssymb}		% allows further math symbols
\usepackage{amsfonts}		% allows further math symbols
\usepackage{mathrsfs}
\usepackage{empheq}
\usepackage{mathtools}		% allows further math commands
\usepackage{bm}             % to make greek letters bold
\usepackage[caption=false,font=footnotesize]{subfig}
\usepackage{graphicx}		% allows figures
\usepackage{booktabs}   	% to improve tables
\usepackage{siunitx}        % to align tables entries to decimal point 
\usepackage[export]{adjustbox}  % to use max width in includegraphics
\usepackage{algorithm}
\usepackage{algpseudocode}

\usepackage{hyperref}		% custom abstract
\newcommand{\T}{^{\mathsf{T}}}
\newcommand{\B}[1]{\if#1\relax\bm{#1}\else\mathbf{#1}\fi} % bold text
\newcommand{\R}[1]{\mathrm{#1}}						      % regular text
\newcommand{\C}[1]{\mathcal{#1}}	
\newcommand{\BB}[1]{\mathbb{#1}}

\graphicspath{{figures/}}  % set images directory
\allowdisplaybreaks
\setlist{noitemsep}       % Remove spacing between bullet/numbered list elements
\sethlcolor{LightGrey}
\hypersetup{			  % set hyperlinks color
    colorlinks,
    linkcolor={MediumBlue},
    citecolor={MediumBlue},
    urlcolor={MediumBlue}
}
\begin{document}

\begin{frontmatter}
%\runtitle{Insert a suggested running title}  % Running title for regular papers but only if the title is over 5 words. Running title is not shown in output.

\title{{\color{black}Convergence and Synchronization in Networks of Piecewise-Smooth Systems via Distributed Discontinuous Coupling}}%\thanksref{footnoteinfo}} % Title, preferably not more than 10 words.

%\thanks[footnoteinfo]{This paper was not presented at any IFAC meeting. Corresponding author M.~T.~Cicero. Tel. +XXXIX-VI-mmmxxi. Fax +XXXIX-VI-mmmxxv.}

\author[Naples]{Marco Coraggio}\ead{marco.coraggio@unina.it},
\author[Naples]{Pietro DeLellis}\ead{pietro.delellis@unina.it}, 
\author[Naples,Bristol]{Mario di Bernardo}\ead{mario.dibernardo@unina.it}

\address[Naples]{Department of Electrical Engineering and Information Technology, University of Naples Federico II, Via Claudio 21, 80125 Naples, Italy}
\address[Bristol]{Department of Engineering Mathematics, University of Bristol, Woodland Road, Clifton BS8 1UB, Bristol, U.K.}

% Five to ten keywords, chosen from the IFAC keyword list or with the help of the Automatica keyword wizard.

\begin{abstract}     % no more than 200 words!
{\color{black}Complex networks are a successful framework to describe collective behaviour in many applications, but a notable gap remains in the current literature, that of proving asymptotic convergence in networks of piecewise-smooth systems.
Indeed, a wide variety of physical systems display discontinuous dynamics that change abruptly, including dry friction mechanical oscillators, electrical power converters, and biological neurons.
%Nonetheless, to this day it has never been shown that a distributed interaction between piecewise-smooth agents can be used to induce convergence of the nodes' trajectories.
%Such collective behaviour has applicability in designing efficient multi-component rotating mechanical parts, in determining frequency consensus in power grids, and more.
In this paper, we study how to enforce global asymptotic state-synchronization in these networks.
Specifically, we propose the addition of a distributed discontinuous coupling action to the commonly used diffusive coupling protocol.
Moreover, we provide analytical estimates of the thresholds on the coupling gains required for convergence, and highlight the importance of a new connectivity measure, which we named \emph{minimum density}.
The theoretical results are illustrated by a set of representative examples.}
\end{abstract}

\end{frontmatter}

% CONTENT %%%%%%%%%%%%%%%%%%%%%%%%%%%%%%%%%%%%%%%%%%%%%%%%%%%%%%%%%%%%

%---------------------------------------------------------------------
\section{Introduction}
\label{sec:introduction}

\color{black}
Complex networks have proven to be a powerful and effective paradigm to describe interaction and collective behaviour in networks of dynamical systems \cite{arenas2008synchronization}.
%\cite{strogatz2001exploring}
Yet, in the vast literature on synchronization of complex networks of the past 20 years, there is a notable gap:
%Consider the following examples: a series of interconnected faults in earthquake modelling \cite{burridge1967model}, biological tissues made of cardiac \cite{berger2007perioddoubling, dibernardo2008piecewisesmooth} or neuronal cells \cite{coombes2016synchrony, coombes2018networks}, the driveline in a vehicle \cite{leine2004dynamics}, power-electronic devices in electrical grids \cite{tse2004complex}, and intertwined gene regulatory networks \cite{casey2006piecewiselinear}.
the case where the agents display discontinuous behaviour and therefore are modelled as \emph{piecewise-smooth} (\emph{PWS}) \emph{dynamical systems} \cite{dibernardo2008piecewisesmooth, cortes2008discontinuous, liberzon2012switching}.
Importantly, differential equations with discontinuous right-hand sides appear in all applications featuring control systems with bang-bang, switched or hybrid controllers \cite{tsypkin1984relay}.
However, \emph{assessing whether coupling between agents will trigger synchronization in a network where the nodes are piecewise-smooth systems is still an open question}.
This is of crucial importance, for example, when the frequencies of multiple power generators must be synchronized in a smart grid with switching components \cite{dorfler2016breaking,tse2004complex}, or when ensuring convergence of the velocities of several interconnected mechanical components subject to dry friction \cite{galvanetto1999nonlinear,leine2004dynamics}.
At the same time, spike synchronization in neurons has a crucial role in activities such as vision and motor coordination \cite{engel1999temporal,coombes2016synchrony, coombes2018networks}, while  convergence of the motion of seismic faults leads the dreadful scenario of an earthquake \cite{scholz2010large,burridge1967model}.

The reason why proving converge in a network of piecewise-smooth systems is a difficult task is twofold. 
Firstly, many common mathematical tools used when proving synchronization (e.g. Lyapunov approaches or the \emph{master stability function} (\emph{MSF}) technique \cite{pecora1998master}), in their standard form, require some degree of differentiability in the agents' vector fields.
Secondly, assumptions on the vector fields typically made for smooth systems, such as one-sided Lipschitz continuity and \emph{QUADness} \cite{delellis2011quad}, cannot be exploited for PWS systems, but for a number of special cases.
This fact also entails that, as discussed in \cite{liu2011dissipativity}, a linear diffusive coupling protocol (that is the most common communication protocol in complex networks) is not sufficient to guarantee convergence.

Numerous studies attempted to find alternative solutions to this problem.
Early results on local synchronization between two coupled PWS systems were illustrated in \cite{danca2002synchronization}, while the specific case of friction oscillators was discussed in \cite{galvanetto1999nonlinear}.
More recently, an extension of the MSF technique to assess local stability of the synchronization manifold was presented in \cite{coombes2016synchrony}, even though, as explained in \cite{lai2018analysis}, such technique is only feasible for \emph{piecewise-smooth continuous} \cite{dibernardo2008piecewisesmooth} systems.
Also, bounded synchronization between two coupled neural networks was investigated in \cite{liu2011dissipativity}, whereas more generic sufficient conditions for global bounded convergence to a synchronous solution were later given in \cite{delellis2015convergence}.
Asymptotic synchronization was achieved via a leader-follower strategy in \cite{yang2013finitetime}, where the authors exploited a control law injected at all the nodes to make the network track the state of a leader agent.
However, this approach requires the presence of a single agent that can exert a centralised control action on all the systems (which is relatively costly and sometimes unfeasible) and the assumption that the trajectories of the open-loop network are bounded.
So far, the only attempts at finding conditions to ensure global asymptotic convergence via a distributed coupling strategy (rather than a centralised controller) are illustrated in \cite{liu2012new, coraggio2018synchronization}.
In both papers strict assumptions (e.g. QUADness and \emph{semi-QUADness} \cite{liu2012new}) are made on the vector fields of the agents, excluding many PWS systems that do not satisfy them \cite{delellis2015convergence,yang2013finitetime,liu2012filippov}.

In this paper, we aim at finding sufficient conditions on the agent dynamics, the communication protocol and the structure of the interconnections in a network of piecewise-smooth systems so as to ensure global asymptotic convergence.
We build on an intuition we presented in \cite{coraggio2018synchronization}, where only a preliminary numerical investigation was carried out.
Specifically, assuming the presence of a linear diffusive coupling protocol in the network, we propose a multiplex control approach \cite{burbanolombana2016multiplex}, where communication among the nodes is extended via an additional discontinuous coupling layer, whose topology might differ from that of the diffusive one.
Such an approach is corroborated by the evidence in the literature that some kind of discontinuous action is often required to enforce global convergence \cite{liu2011dissipativity, yang2013finitetime}.
%We argue that the cheapest and most natural way to implement such an action  is injecting it in the coupling law in a distributed fashion.
In addition, discontinuous coupling protocols were already effectively used in different contexts. 
For example, they were exploited in \cite{cortes2006finitetime,hui2010finitetime}, to drive networks of integrators to consensus in finite time, and in \cite{mayhew2012quaternionbased, wei2018finitetime}, using a hybrid system framework, to solve the problem of attitude synchronization among quaternions with smooth dynamics.

With the above multiplex distributed strategy, we advance the current state-of-the-art as follows.
\begin{enumerate}
\item We prove formally for the first time that asymptotic (rather than just bounded) synchronization in networks of piecewise-smooth systems is possible through the use of a distributed, multiplex coupling strategy;
%, and without resorting to leader-follower strategies, that are costly or sometimes infeasible;
\item We give analytical estimates of the critical values of both the coupling gains associated to the diffusive and the discontinuous layers, sufficient to guarantee global synchronization;
\item We show that the threshold value for the discontinuous coupling gain is dependent on a quantity we name {\em minimum density} of a graph, that acts as a connectivity measure, similarly to the well-known algebraic connectivity \cite{godsil2013algebraic}.
\end{enumerate}
\color{black}
%All the theoretical results are illustrated via numerical simulations on a set of representative examples, highlighting the advantages and disadvantages of the proposed approach.

%---------------------------------------------------------------------
\section{Notation and description of the problem}
\label{sec:description_of_the_problem}

%---------------------------------------------------------------------
\subsection{Notation}

%$\triangleq$ means ``is defined as''.
%Given a scalar $\alpha$, we denote by $\left\lvert \alpha \right\rvert$ its absolute value; by
%$\R{sign}(\alpha)$ the sign of $\alpha$ ($\R{sign}(0) = 0$); by
%$\lfloor \alpha \rfloor$  the largest integer $\beta$ such that $\beta \le \alpha$, while $\lceil \alpha \rceil$ is the smallest integer $\beta$ such that $\beta \ge \alpha$.
%
Given a vector $\B{\xi} = [ \xi_1 \ \xi_2 \ \cdots \ \xi_n ]$, $\left\lvert \B{\xi} \right\rvert = [ \left\lvert \xi_1 \right\rvert \ \left\lvert \xi_2 \right\rvert \ \cdots \ \left\lvert \xi_n \right\rvert ]$;
$\R{sign}(\B{\xi}) = [ \R{sign}(\xi_1) \ \R{sign}(\xi_2) \ \cdots \ \R{sign}(\xi_n) ]$;
$\B{i}_i \in \BB{R}^{n}$ is the column vector with $1$ in position $i$ and $0$ elsewhere;
$\R{diag}(\B{\xi})$ is the diagonal matrix having the elements of vector $\B{\xi}$ on its diagonal.
$\left\lVert \cdot \right\rVert_p$ is the $p$-norm, and we recall that $\left\lVert \B{\xi} \right\rVert_1 \triangleq \sum_{i=1}^n \left\lvert \xi_i \right\rvert$.

Given a set $\C{Q}$, if it is finite, we denote by $\left\lvert \C{Q} \right\rvert$ its cardinality.
%We denote by $\mathscr{P}(\C{Q})$ the \emph{powerset} of $\C{Q}$, that is the set of all subsets of $\C{Q}$.
For a set $\C{Q}$ of scalars, the notation $\C{Q} \le 0$ means that $\forall \alpha \in \C{Q}$, $\alpha \le 0$ (analogously for $\ge$, $=$, etc.).
{\color{black}$\C{Q} \rightrightarrows \C{R}$ denotes an application from $\C{Q}$ to the set of all subsets of $\C{R}$.}
%Finally, we denote by $\BB{R}$ the set of real numbers, by $\BB{N}$ the set of natural numbers including zero, and by $\BB{N}_{>0}$ that which excludes zero.
%Given a function $\phi$, we denote by $\left. \phi\right\rvert_\Omega$ its restriction to the set $\Omega$.

Given a matrix $\B{A}$, {\color{black}we denote by %$\R{sym}(\B{A})$ its symmetric part;
$\lambda_i(\B{A})$ its} $i$-th eigenvalue, with eigenvalues being sorted in an increasing fashion if they are all real (so that $\lambda_{\R{min}}(\B{A}) \triangleq \lambda_1(\B{A})$). The notation
$\B{A} > 0$ indicates that $\B{A}$ is positive definite (analogously for semi- and negative definiteness); 
%$\B{I}_n$ is the $n \times n$ identity matrix, $\B{0}_{n \times m}$ is the $n \times m$ null matrix, and $\B{0}_n$ is the null column vector with $n$ entries; we will omit the subscripts when not necessary.
Finally, $\otimes$ is the Kronecker product.
We recall that $\left\lVert \B{A} \right\rVert_\infty \triangleq {\displaystyle\max_{i = 1, \dots, n}} \left( \sum_{j=1}^{n}\left\lvert A_{ij} \right\rvert \right)$.

%---------------------------------------------------------------------
\subsection{Problem description}

\begin{figure}[t]
    \centering
    \includegraphics[max width=\columnwidth]{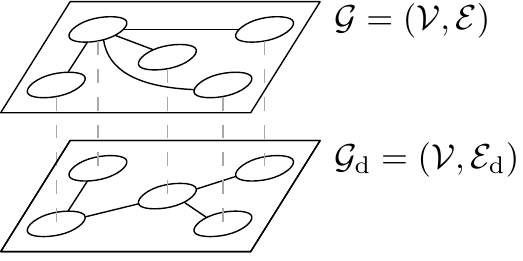}
    \caption{A multilayer network with $N = 5$ and two coupling layers.}
    \label{fig:multilayer_network}
\end{figure}

We consider the problem of finding conditions on the agent vector fields and an appropriate distributed control protocol $\B{u}_i$ that can to make an ensemble of identical PWS systems of the form%
\footnote{\label{ftn:existence_solution}{\color{black}To ensure the existence of a solution, we assume the Filippov vector field defining the systems' dynamics is locally bounded, takes nonempty, compact, and convex values and is upper-semicontinuous; \cite[Propositions S2 and 3]{cortes2008discontinuous}. 
We allow the possibility that solutions are not unique.}}
\begin{equation}\label{eq:network}
\dot{\B{x}}_i = \B{f}(\B{x}_i; t) + \B{u}_i, \quad i = 1, \dots, N
\end{equation}
converge asymptotically towards a common evolution. 
Here $\B{x}_i \in \BB{R}^n$, $t \in \BB{R}$ is time and $\B{f} : \BB{R}^n \times \BB{R} \rightarrow \BB{R}^n$ is a generic piecewise-smooth vector field that might possibly exhibit sliding dynamics \cite{dibernardo2008piecewisesmooth}.

\begin{defn}[Asymptotic synchronization]
Network \eqref{eq:network} is asymptotically synchronized in  $\Omega \subseteq \BB{R}^{nN}$ if, for all initial conditions in $\Omega$,
$\lim_{t \rightarrow +\infty} \left\lVert \B{x}_i(t) - \B{x}_j(t) \right\rVert = 0$,
for all $i,j = 1, \dots, N$.
Moreover, network \eqref{eq:network} is \emph{globally asymptotically synchronized} if $\Omega = \BB{R}^{nN}$.
\end{defn}
%Our goal is to synthesise a distributed feedback control strategy such that network \eqref{eq:network} exhibits global synchronization. 
{\color{black}As shown in \cite{delellis2015convergence}, a purely linear diffusive protocol does not suffice to achieve asymptotic convergence.
Therefore, in this work we propose a multiplex control strategy \cite{burbanolombana2016multiplex} where an additional discontinuous coupling layer is added to a traditional linear diffusive one (see Figure \ref{fig:multilayer_network}).
The two layers are associated to two undirected unweighted graphs: $\C{G} = (\C{V}, \C{E})$ associated to linear diffusive coupling, and $\C{G}_\R{d} = (\C{V}, \C{E}_\R{d})$ associated to discontinuous coupling; here, $\C{V}$ is the set of vertices (or nodes) and $\C{E}, \C{E}_\R{d}$ are the set of edges (or links).}
Thus, our multilayer control action is given by 
\begin{equation}\label{eq:discontinuous_coupling_network}
\B{u}_i = - c \sum_{j=1}^{N} L_{ij} \B{\Gamma} (\B{x}_j - \B{x}_i)- c_\R{d} \sum_{j=1}^{N} L_{ij}^\R{d} \B{\Gamma}_\R{d} \R{sign} (\B{x}_j - \B{x}_i),
\end{equation}
%\begin{equation}\label{eq:discontinuous_coupling_network}
%\begin{split}
%\dot{\B{x}}_i = 
%{}&\B{f}(\B{x}_i; t)
%- c \sum_{j=1}^{N} L_{ij} \B{\Gamma} (\B{x}_j - \B{x}_i)\\
%&- c_\R{d} \sum_{j=1}^{N} L_{ij}^\R{d} \B{\Gamma}_\R{d} \R{sign} (\B{x}_j - \B{x}_i),
%\end{split}
%\end{equation}
for $i = 1, \dots, N$, where $L_{ij}$ and $L_{ij}^\R{d}$ are the $(i,j)$-th elements of the symmetric \emph{Laplacian matrices} $\B{L}, \B{L}_\R{d} \in \BB{R}^{N \times N}$ associated to the graphs $\C{G}$ and $\C{G}_\R{d}$, respectively; 
$\B{\Gamma}, \B{\Gamma}_\R{d} \in \BB{R}^{n \times n}$ are \emph{inner coupling matrices} {\color{black}describing how the coupling actions affect the dynamics of the nodes.}
%Then, the problem becomes that of finding sufficient conditions for global synchronization on the coupling gains $c$ and $c_\R{d}$, the layer topologies $\B{L}$ and $\B{L}_\R{d}$, the node dynamics $\B{f}$, and the inner coupling matrices $\B{\Gamma}$ and $\B{\Gamma}_\R{d}$.
%Before presenting our main convergence results, we expound next, for the sake of clarity, some necessary mathematical preliminaries to make the paper self-contained.

%---------------------------------------------------------------------
\section{Mathematical preliminaries}
\label{sec:notation_and_mathematical_preliminaries}

{\color{black}Let $\B{A} \in \BB{R}^{n \times n}$ be a matrix and $\mu_p : \BB{R}^{n \times n} \rightarrow \BB{R}$ be the \emph{matrix measure} (\emph{logarithimc norm}) induced by the $p$-norm. 
We recall that 
$
\mu_2(\B{A}) = \lambda_{\R{max}} ( \frac{\B{A} + \B{A}\T}{2} )
$
and
$
\mu_\infty(\B{A}) = \max_i ( A_{ii} + \sum_{j = 1, j \ne i}^n \left\lvert A_{ij} \right\rvert ).
$
We denote $- \mu_p(-\B{A})$ by $\mu_p^-(\B{A})$, so that
$
\mu^-_2(\B{A}) = \lambda_{\R{min}} (\frac{\B{A} + \B{A}\T}{2}),
$
and
$
\mu_\infty^-(\B{A}) = \min_{i} ( A_{ii} - \sum_{j = 1, j \ne i}^n \left\lvert A_{ij} \right\rvert ).
$
}
We also note that $\mu_{\infty}^-(\B{I}_m \otimes \B{A}) = \mu_\infty^-(\B{A})$, for any $m \in \BB{N}_{>0}$.

\subsection{\textsigma-QUAD property}
\label{sec:sigma_quad_property}

We define next an extension to the well-known QUAD assumption, often used as a regularity condition on the internal agent dynamics in synchronization problems \cite{delellis2011quad, coraggio2018synchronization}.

\begin{figure}[t]
    \centering
    \includegraphics[max width=\columnwidth]{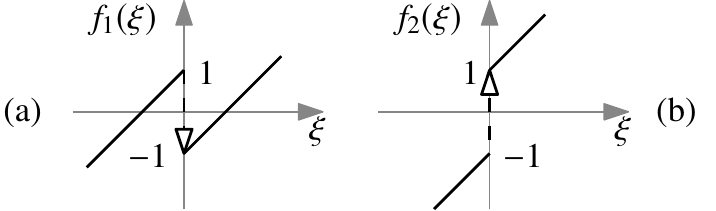}
    \caption{(a) $f_1$ is QUAD and \textsigma-QUAD; (b) $f_2$ is not QUAD, but is \textsigma-QUAD.}
    \label{fig:quad_nonquad_v02}
\end{figure}

\begin{defn}[\textsigma-QUAD]\label{def:sigma_quad}
A function $\B{f} : \BB{R}^n \times \BB{R} \rightarrow \BB{R}^n$ is \textsigma-QUAD($\B{P}$, $\B{Q}$, $\B{M}$) if, $\forall \B{\xi}_1, \B{\xi}_2 \in \BB{R}^n$, $t \in \BB{R}$, $\exists \B{P}, \B{Q}, \B{M} \in \BB{R}^{n\times n}$ such that
\begin{equation*}%\label{eq:sigma_quad}
\begin{aligned}
&\left( \B{\xi}_1\hspace{-0.01cm} - \B{\xi}_2 \right)\T\hspace{-0.05cm}\B{P}\hspace{-0.01cm} \left[ \B{f}(\B{\xi}_1; t)\hspace{-0.01cm} - \B{f}(\B{\xi}_2; t) \right]\hspace{-0.01cm} \le \hspace{-0.01cm}
\left( \B{\xi}_1 - \B{\xi}_2 \right)\T\hspace{-0.05cm} \B{Q}\hspace{-0.01cm} \left( \B{\xi}_1\hspace{-0.01cm} - \B{\xi}_2 \right) \\
&+ \left( \B{\xi}_1\hspace{-0.01cm} - \B{\xi}_2 \right)\T\hspace{-0.05cm} \B{M} \ \R{sign} \left( \B{\xi}_1\hspace{-0.01cm} - \B{\xi}_2 \right).
\end{aligned}
\end{equation*}
\end{defn}
{\color{black}Note that if $\B{M} = \B{0}_{n \times n}$, Definition \ref{def:sigma_quad} becomes equivalent to the classic QUAD condition.}
However, differently from the QUAD condition, the \textsigma-QUAD property includes cases where $\B{f}$ has any number of finite jumps discontinuities.
{\color{black}Indeed, many real-world systems satisfy the \textsigma-QUAD property (while failing to verify the QUAD assumption), featuring neuron models \cite{coombes2016synchrony}, gene regulatory networks \cite{casey2006piecewiselinear}, DC-DC converters, dry friction mechanical oscillators \cite{galvanetto1999nonlinear}, and all QUAD systems controlled with bang-bang controllers \cite{dibernardo2008piecewisesmooth}.
}
%A related concept is the QUAD-affine condition defined in \cite{delellis2015convergence}, where an affine term is added to the quadratic term in \eqref{eq:sigma_quad} to account for the presence of finite jumps.
%Moreover, in \cite{liu2012new}, the notion of a non-autonomous vector field being \emph{semi-QUAD} is introduced. 
%Specifically, a vector field, say $\B{f}(\B{\xi};t)$, is said to be semi-QUAD if its difference with another vector field $\B{g}(\B{\xi};t)$, known to be QUAD, tends asymptotically to zero as time increases. 
%Finally, another related concept is the \emph{growth condition} for a vector field, defined in \cite{liu2012filippov}.
As an illustrative example of what the \textsigma-QUAD property implies, consider the functions $f_1, f_2 : \BB{R} \rightarrow \BB{R}$, given by $f_1 (\xi) = \xi - \R{sign}(\xi)$ and $f_2(\xi) = \xi + \R{sign}(\xi)$, represented in Figure \ref{fig:quad_nonquad_v02}.
$f_1$ is \textsigma-QUAD with $\B{P} = 1$, $\B{Q} = 1$, $\B{M} = 0$, and thus is also QUAD; differently, $f_2$ is \textsigma-QUAD with $\B{P} = 1$, $\B{Q} = 1$, $\B{M} = 2$, and can be proved not to be QUAD.
%As a further note, we remark that QUADness and \textsigma-QUADness are more general than passivity, which has also been used to prove synchronization in complex networks [REF?].
%For example, it is straightforward to verify that passive systems are globally stable in the origin (in the absence of an external input), which is not necessarily the case for QUAD systems.%
%Besides, the traditional definition of passivity is not applicable to discontinuous functions, because the storage function would not be continuously differentiable.

{\color{black}The following Lemma can be used to check for \textsigma-QUADness. 
The proof immediately follows from the definitions of QUADness and \textsigma-QUADness.
\begin{lem}[Evaluation of \textsigma-QUADness]\label{lem:evaluate_sigma_quad}
Given a function $\B{f} : \BB{R}^n \times \BB{R} \rightarrow \BB{R}^n$, if there exist $\B{f}_\R{Q}$ and $\B{f}_\R{\sigma}$ such that 
(i) $\B{f} = \B{f}_\R{Q} + \B{f}_\R{\sigma}$, 
(ii) $\B{f}_\R{Q}$ is QUAD($\B{P}$, $\B{Q}$), and 
(iii) $\B{f}_\sigma$ is bounded---i.e. $\exists \B{m} \in \BB{R}^n : \left\lvert \B{f}_\sigma(\B{\xi}_1) - \B{f}_\sigma(\B{\xi}_2) \right\rvert \le \B{m}$, for all $\B{\xi}_1, \B{\xi}_2 \in \BB{R}^n$---then $\B{f}$ is \textsigma-QUAD($\B{P}$, $\B{Q}$, $\B{M}$) with $\B{M} = \R{diag}(\left\lvert \B{P} \right\rvert \B{m})$.
\end{lem}}
% NB non ho inserito la proof per risparmiare spazio
{\color{black}
\begin{rem}\label{rem:evaluate_sigma_quad}
Lemma \ref{lem:evaluate_sigma_quad} allows to assess whether a PWS system with any number of finite jumps is \textsigma-QUAD by only looking at its Jacobian, where it is defined. Indeed, if a vector field has a bounded Jacobian, then it also fulfils the QUAD assumption
\cite[Proposition 12]{coraggio2020distributed}.%
\footnote{{\color{black}\label{ftn:quad_systems}As a matter of fact, many well-known systems are QUAD, e.g.~see \cite{delellis2011quad}.}}
\end{rem}
}

%---------------------------------------------------------------------
\subsection{Network notation and definitions}
\label{sec:network_notation_and_definitions}

With reference to network \eqref{eq:network}, \eqref{eq:discontinuous_coupling_network}, $\tilde{\B{x}} \triangleq \sum_{i=1}^{N} \B{x}_i/N \in \BB{R}^n$ is the \emph{average of the states} of the nodes;
$\B{e}_i \triangleq \B{x}_i - \tilde{\B{x}} \in \BB{R}^n$, $i = 1, \dots, N$ are the \emph{synchronization errors}, and we denote the $h$-th element of $\B{e}_i$ by $e_{i,h}$;
$\bar{\B{x}} \triangleq [ \B{x}_1\T \ \cdots \ \B{x}_N\T ]\T \in \BB{R}^{nN}$ is the \emph{stack of the states} of the nodes;
$\bar{\B{e}} \triangleq [ \B{e}_1\T \ \cdots \ \B{e}_N\T ]\T \in \BB{R}^{nN}$ is the \emph{stack of the errors};
$\B{e}^h \triangleq [ e_{1,h} \ \cdots \ e_{N,h} ]\T \in \BB{R}^N$ groups the $h$-th components of all the {\color{black}errors $\B{e}_i$};
$e_\R{tot} \triangleq \frac{1}{N} \sum_{i=1}^{N} \left\lVert \B{e}_i \right\rVert_2$ is the \emph{total synchronization error}.

Given a graph $\C{G}=\left\{\C{V}, \C{E} \right\}$, we denote by $N$ the number of its vertices and by $N_\C{E}$ be the number of its edges.
%For instance, in a network with $N=5$ and $N_\C{E} = 3$, the existence of edge number 3, connecting vertices $v_2$ and $v_4$, would imply that $\B{b}_3 = [ 0 \ 1 \ 0 \ - \! 1 \ 0 ]\T$, or equivalently $\B{b}_3 = [ 0 \ - \! 1 \ 0 \ 1 \ 0 ]\T$.
A \emph{cut} $\C{C}$ is a partition of the set of vertices $\C{V}$ in two subsets $\C{V}_1$, $\C{V}_2$; we name the sets of all possible cuts on $\C{G}$ as $\hat{\C{C}}_{\C{G}}$, the number of edges that connect a vertex in $\C{V}_1$ with one in $\C{V}_2$ as $b$, and define the cardinalities $N_1 = \left\lvert \C{V}_1 \right\rvert$, $N_2 = \left\lvert \C{V}_2 \right\rvert$.
In what follows, the ratio $b/N_1 N_2$ represents the \emph{density} of a cut $\C{C}$.

\begin{defn}[Minimum density]\label{def:minimum_density}
The \emph{minimum density} of a graph $\C{G}$ is given by
    \begin{equation}\label{eq:minimum_density}
    \delta_{\C{G}} \triangleq \frac{N}{2} \min_{\C{C} \in \hat{\C{C}}_\C{G}} {\frac{b}{N_1 N_2}}.
    \end{equation}
\end{defn}
The optimal cut $\arg \min_{\C{C} \in \hat{\C{C}}_\C{G}} \frac{b}{N_1 N_2}$ associated to $\delta_{\C{G}}$ is known as the \emph{sparsest cut} \cite{matula1990sparsest}, and is here denoted by $\C{C}_{\text{sc}}$.
The problem of finding such a cut is called the \emph{sparsest cut problem}, which is a special kind of \emph{graph partitioning} problem. %\cite{buluc2016recent}

{\color{black}The sparsest cut problem is NP-hard and is normally solved algorithmically, as discussed below and in \cite{arora2010sqrt}. 
}%arora2008geometry
{\color{black}The minimum density of a graph can be computed using the free software \textsc{METIS} \cite{karypis1998fast}, which partitions the vertices into two subsets, minimising the number of edges between them, while keeping the sizes of the subsets to some fixed values.
Specifically, we run \textsc{METIS} $\lfloor N/2 \rfloor$ times, each time constraining $(N_1, N_2)$ to be $(1, N-1)$, then $(2, N-2)$ and so on until $(\lfloor N/2 \rfloor, \lceil N/2 \rceil)$.
At each run, we compute $\frac{N}{2} \frac{b}{N_1 N_2}$ and eventually choose the smallest value as the minimum density of the graph, according to Definition \ref{def:minimum_density}.}

Note that the minimum density of some selected graph topologies can be computed analytically by relatively simple algebra. 
We report these findings in Table \ref{tab:topologies} (the proofs are omitted here for the sake of brevity).
%\begin{itemize}
%    \item for a complete graph, $\delta = N/2$;
%    \item for a path graph, $\delta = 2/N$ if $N$ is even, and $\delta = 2N/(N^2-1)$ if $N$ is odd;
%    \item for a ring graph, $\delta = 4/N$ if $N$ is even, and $\delta = 4N/(N^2-1)$ if $N$ is odd;
%    \item for a star graph, $\delta = N/(2(N-1))$;
%    \item for a nearest-neighbour graph, $\delta = ( 4 \sum_{k = 0}^{l-1} (l-k) ) / N$ if $N$ is even, and $\delta = ( 4N \sum_{k = 0}^{l-1} (l-k) ) / (N^2-1)$ if $N$ is odd.
%\end{itemize}

\begin{table}[t]
    \caption{Values of the minimum density $\delta_{\C{G}}$, the algebraic connectivity $\lambda_2(\B{L})$ \cite{fiedler1990absolute}, and the number of edges $N_{\C{E}}$ for selected topologies. ``$l$-near.-n.'' stands for ``$l$-nearest-neighbours''; $\R{c_1} \triangleq \cos{\left(\pi/N\right)}$, $\R{c_2} \triangleq \cos{\left(2\pi/N\right)}$.}%The values of the algebraic connectivity are taken from\cite{fiedler1990absolute} .}
    \label{tab:topologies}
    \begin{center}
        \begin{tabular}{@{}l|lll@{}}
            \toprule
            Topology & $\phantom{\begin{dcases}\end{dcases}}\delta_{\C{G}}$ & $\lambda_2(\B{L})$ & $N_{\C{E}}$ \\ \midrule
            Complete & $\phantom{\begin{dcases}\end{dcases}}N/2$ & $N$ & $\frac{N^2-N}{2}$\\
            Star & $\phantom{\begin{dcases}\end{dcases}}N/(2(N-1))$ & $1$ & $N-1$\\
            Path & $\begin{cases} 2/N, & \text{$N$ even} \\ 2N/(N^2-1), & \text{$N$ odd} \end{cases}$ & $2\left(1-\R{c}_1\right)$ & $N-1$\\
            Ring & $\begin{cases} 4/N, & \text{$N$ even} \\ 4N/(N^2-1), & \text{$N$ odd} \end{cases}$ & $2\left(1-\R{c}_2\right)$ & $N$ \\
            $l$-near.-n. & $\begin{cases} \frac{4 \sum_{k = 0}^{l-1} (l-k)}{N}, & \text{$N$ even} \\ \frac{4N \sum_{k = 0}^{l-1} (l-k)}{N^2-1}, & \text{$N$ odd} \end{cases}$ & - & $Nl$\\
            \bottomrule
        \end{tabular}
    \end{center}
\end{table}

%---------------------------------------------------------------------
\section{Convergence results}
\label{sec:convergence_results}

We expound next our main convergence results that allow to estimate the value of the critical coupling gains $c^*$ and $c_\R{d}^*$ of the diffusive and discontinuous coupling layers, respectively, that guarantee global convergence of all agents towards a common synchronous evolution. 

\begin{thm}\label{thm:discontinuous_coupling}
Network \eqref{eq:network} controlled by the distributed multiplex control action \eqref{eq:discontinuous_coupling_network} achieves global asymptotic synchronization if
\begin{enumerate}[(a)]
\item there exist $\B{P}, \B{Q}, \B{M} \in \BB{R}^{n \times n}$, with $\B{P} > 0$, such that (i) $\B{f}$ is \textsigma-QUAD($\B{P}$, $\B{Q}$, $\B{M}$), (ii) {\color{black}$\mu_2^-(\B{P}\B{\Gamma}) > 0$}, and $\mu_\infty^-(\B{P}\B{\Gamma}_\R{d}) > 0$, %$\R{sym}(\B{P} \B{\Gamma}) > 0$
\item $\C{G}$ and $\C{G}_\R{d}$ are connected graphs, and
\item $c>c^*$, $c_\R{d} \ge c_\R{d}^*$ with   
\begin{equation}\label{eq:thresholds}
c^* \triangleq \frac{{\color{black}\mu_2(\B{Q})}} {\lambda_{2} (\B{L}) \ \ {\color{black}\mu_2^-(\B{P}\B{\Gamma})} },  % \lambda_{\R{min}} (\R{sym} \ \B{P}\B{\Gamma} )%
\quad
c^*_\R{d} \triangleq \frac{{\color{black}\mu_\infty(\B{M})}}{\delta_{\C{G}_\R{d}} \ \ \mu_\infty^-(\B{P}\B{\Gamma}_\R{d})}.
\end{equation}    
\end{enumerate}
\end{thm}

A proof of this theorem is given later in Section \ref{sec:proofs}. Here, we wish to emphasise that the critical coupling gains {\color{black}$c^*, c^*_\R{d}$} depend on the internal node dynamics $\B{f}$ through the matrices $\B{Q}$, $\B{P}$, $\B{M}$, the inner coupling matrices $\B{\Gamma}$, $\B{\Gamma}_\R{d}$, and the structure of the control layers $\B{L}$, $\B{L}_\R{d}$ via the algebraic connectivity $\lambda_2(\B{L})$ and the minimum {\color{black}density $\delta_{\C{G}_\R{d}}$}.
%Hence, the convergence theorem above can be effectively used to design the network control layers as illustrated via representative examples in Section \ref{sec:numerical_example}.
%Note {\color{black}that Theorems \ref{thm:discontinuous_coupling} gives} sufficient conditions on the threshold values of the coupling gain that scale with $\lambda_2(\B{L})^{-1}$ for $c^*$ and $\delta_{\C{G}_\R{d}}^{-1}$ for $c_\R{d}^*$. 
%For the sake of completeness, Table \ref{tab:topologies} shows how these structural variables change for a set of paradigmatic network topologies of $N$ vertices together with their total number of edges.
The multiplex nature of the strategy proposed here allows to pick the structure of each layer so as to fulfil a trade-off between the values of the required coupling gains and the number of edges in each layer (for some selected topologies, see Table \ref{tab:topologies}).

%Next, we provide an alternative condition for global synchronization to deal with the case in which the inner coupling matrices $\B{\Gamma}$ and $\B{\Gamma}_\R{d}$ do not fulfil the conditions $\R{sym}(\B{P} \B{\Gamma}) > 0$ and $\mu_\infty^-(\B{P}\B{\Gamma}_\R{d}) > 0$ in Theorem \ref{thm:discontinuous_coupling}.

{\color{black}
\begin{rem}\label{rem:smooth_networks}
The results in Theorem \ref{thm:discontinuous_coupling} encompass the case of networks of smooth systems, whose global synchronization was characterised in \cite{delellis2011quad}.
As a matter of fact, in that case the systems would be \textsigma-QUAD with $\B{M} = \B{0}$ (i.e.~QUAD) and \eqref{eq:thresholds} would yield $c_\R{d}^* = 0$ and $c^*$ corresponding to \cite[(H2)]{delellis2011quad}.
We also note that Theorem \ref{thm:discontinuous_coupling} accounts for the possible occurrence of sliding solutions, as explained in Section \ref{sec:semi_negativity_of_W_2}.
\end{rem}
}

%%---------------------------------------------------------------------
%\subsection{Evaluation of the conditions in Theorem \ref{thm:discontinuous_coupling}}

{\color{black} 
Concerning the evaluation of the conditions in Theorem \ref{thm:discontinuous_coupling}, one possible way to assess \textsigma-QUADness is through Lemma \ref{lem:evaluate_sigma_quad} and \cite[Proposition 12]{coraggio2020distributed}.
Then, to estimate the algebraic connectivity $\lambda_2(\B{L})$ it is possible to use methods such as that illustrated in \cite{yang2010decentralized}.
As far as $\delta_{\C{G}_\R{d}}$ is concerned, to the best of our knowledge, no techniques to estimate it locally have been devised yet, which is reasonable though, since this is the first time this quantity is used in the context of network synchronization.
Hence, until new estimation techniques are developed, it is advisable to compute $\delta_{\C{G}_\R{d}}$ through the algorithm or the formulae we provided in Section \ref{sec:network_notation_and_definitions}, and transmit this information to the nodes directly as a one-time operation.
In addition, hypothesis (ii) holds if $\B{P} \B{\Gamma} > 0$ and $\B{P} \B{\Gamma}_\R{d} > 0$, which in turn hold if 
$\B{\Gamma} > 0$,
$\B{P} \B{\Gamma} = \B{\Gamma} \B{P}$, and
$\B{\Gamma}_\R{d} > 0$,
$\B{P} \B{\Gamma}_\R{d} = \B{\Gamma}_\R{d} \B{P}$, which is as restrictive as other assumptions made in the literature on global synchronization (e.g. \cite{yang2013finitetime,delellis2011quad,liu2012new,delellis2015convergence}).%mishra2018robust, cheng2018aperiodically, feng2017finitetime, zhang2013exponential
%We make a comment to the cases in which $\B{\Gamma}$ and $\B{\Gamma}_\R{d}$ are not positive definite in \ref{item:r2_Gamma}.
}

\section{Examples}
\label{sec:numerical_example}

%---------------------------------------------------------------------
\subsection{Achieving synchronization}

We consider the problem of achieving global asymptotic synchronization in a network of $N=30$ piecewise-smooth relay systems {\color{black}with} $\B{f}(\B{x}_i) = \B{A} \B{x}_i - \B{B} \R{sign}(x_{i,1})$, {\color{black}where}
$\B{A} = \left[ \begin{smallmatrix}
1.51    & 1 & 0 \\
-99.922 & 0 & 1 \\
-5      & 0 & 0
\end{smallmatrix} \right]$
and
$\B{B} = [1 \ -\! 2 \ 1]\T$.
{\color{black}Such systems naturally exhibit chaotic behaviour \cite{dibernardo2008piecewisesmooth} and would not synchronize unless appropriately coupled.}
%In \cite{delellis2015convergence}, it is shown that under some hypotheses a network of such relays can achieve bounded convergence to the synchronous manifold. 
%We show below that using Theorem \ref{thm:discontinuous_coupling}, we can prove instead global asymptotic convergence.
{\color{black}The internal dynamics} $\B{f}$ can be shown to be \textsigma-QUAD according to Definition \ref{def:sigma_quad} through simple algebraic manipulations, with $\B{P} = \B{I}_3$, $\B{Q} = \B{A}$, and
$\B{M} = \left[ \begin{smallmatrix}
0 & 0 & 0 \\ 
4 & 0 & 0 \\ 
0 & 0 & 0
\end{smallmatrix} \right]$; therefore ${\color{black}\mu_2(\B{Q}) = 50.312}$ and ${\color{black}\mu_\infty(\B{M})} = 4$.
We assume that the relays are coupled via two layers as in \eqref{eq:discontinuous_coupling_network}, with the structure of the {\color{black}diffusive} layer, $\B{L}$, corresponding to a ring graph, with $\lambda_2(\B{L}) = 1$, while the structure of the discontinuous coupling layer, $\B{L}_\R{d}$, being chosen as the {\color{black}Erd\H{o}s-R\'enyi-like graph shown} in Figure \ref{fig:L_d_random_N_30_a}, {\color{black}with minimum density }$\delta_{\C{G}_\R{d}} = 1.290$.
%\cite{erdos1960evolution}
Figure \ref{fig:L_d_random_N_30_b} shows the sparsest cut of this latter graph, obtained numerically.
We assume all states are available for coupling so that $\B{\Gamma} = \B{\Gamma}_\R{d} = \B{I}_3$; hence, ${\color{black}\mu_2^-(\B{P}\B{\Gamma})} = \mu_\infty^-(\B{P}\B{\Gamma}_\R{d}) = 1$.
From Theorem \ref{thm:discontinuous_coupling}, we can then compute the critical coupling gains that are sufficient for global convergence as ${\color{black}c^* = 50.312}$ and $c_\R{d}^* = 3.102$.

\begin{figure}[t]
    \centering
%    \subfloat[]{\includegraphics[max width=\columnwidth]{c_01_cd_0001_ring_step_0001_time.pdf}
%    \label{fig:simulation_thresholds_a}}\\
    \subfloat[]{\includegraphics[scale=0.8]{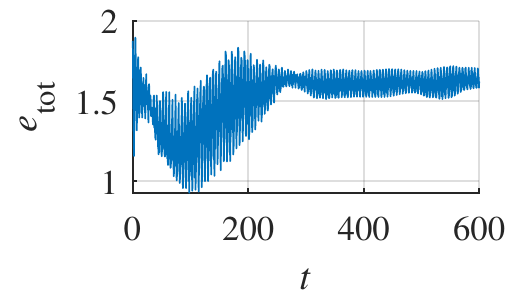}
    \label{fig:simulation_thresholds_b}}
    \subfloat[]{\includegraphics[scale=0.8]{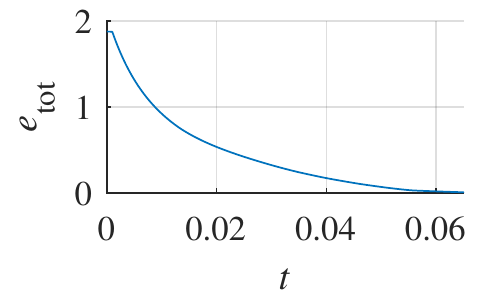}
    \label{fig:simulation_thresholds_c}}
    \caption{Error dynamics in a network of relay systems with (a) $c = 0.1$, $c_\R{d} = 0.001$ and with (b) $c = {\color{black}51}$, $c_\R{d} = 3.200$.}
    \label{fig:simulation_thresholds}
\end{figure}

Figures \ref{fig:simulation_thresholds_b} and \ref{fig:simulation_thresholds_c} show the evolution of the total synchronization error $e_\R{tot}$ (defined in Section \ref{sec:notation_and_mathematical_preliminaries}) when the coupling gains are chosen below and above the critical threshold values.
As expected from the theoretical results, when the gains are above the thresholds, the synchronization error converges asymptotically to zero.
Note that the analytical estimates of the critical coupling gains are very conservative, as expected from a Lyapunov-based proof of convergence.

%---------------------------------------------------------------------
\subsection{Assessing network resilence}

\begin{figure}[t]
    \centering
    \subfloat[]{\includegraphics[max width=0.49\columnwidth]{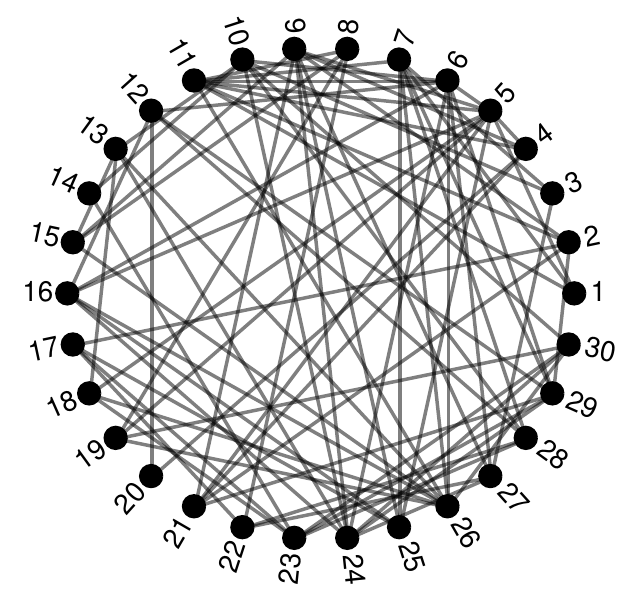}
    \label{fig:L_d_random_N_30_a}}
    \subfloat[]{\includegraphics[max width=0.49\columnwidth]{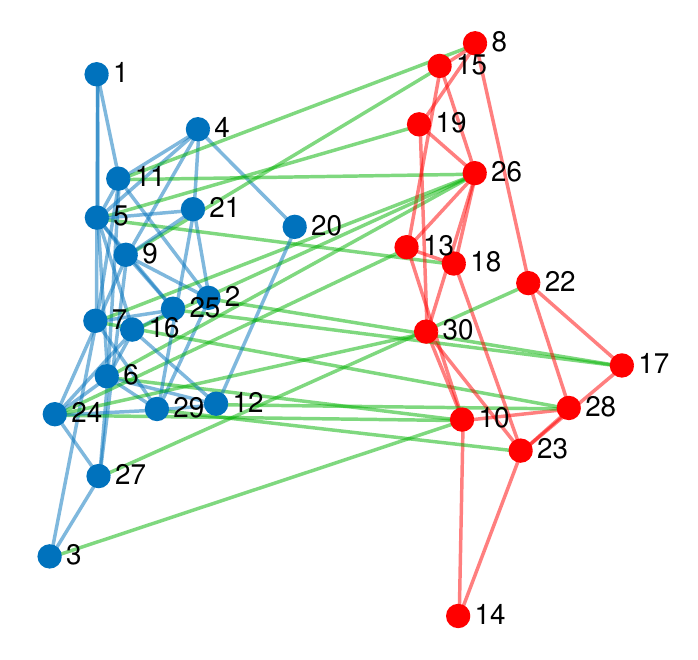}
    \label{fig:L_d_random_N_30_b}}
    \caption{(a) Topology of the discontinuous coupling layer. The graph is an Erd\H{o}s-R\'enyi-like one, with probability parameter $p=0.2$. (b) Sparsest cut of the topology in (a); $N_1 = 17$, $N_2 = 13$. $b = 19$.}
    \label{fig:L_d_random_N_30}
\end{figure}

\begin{figure}[t]
    \centering
    \subfloat[]{\includegraphics[max width=0.49\columnwidth]{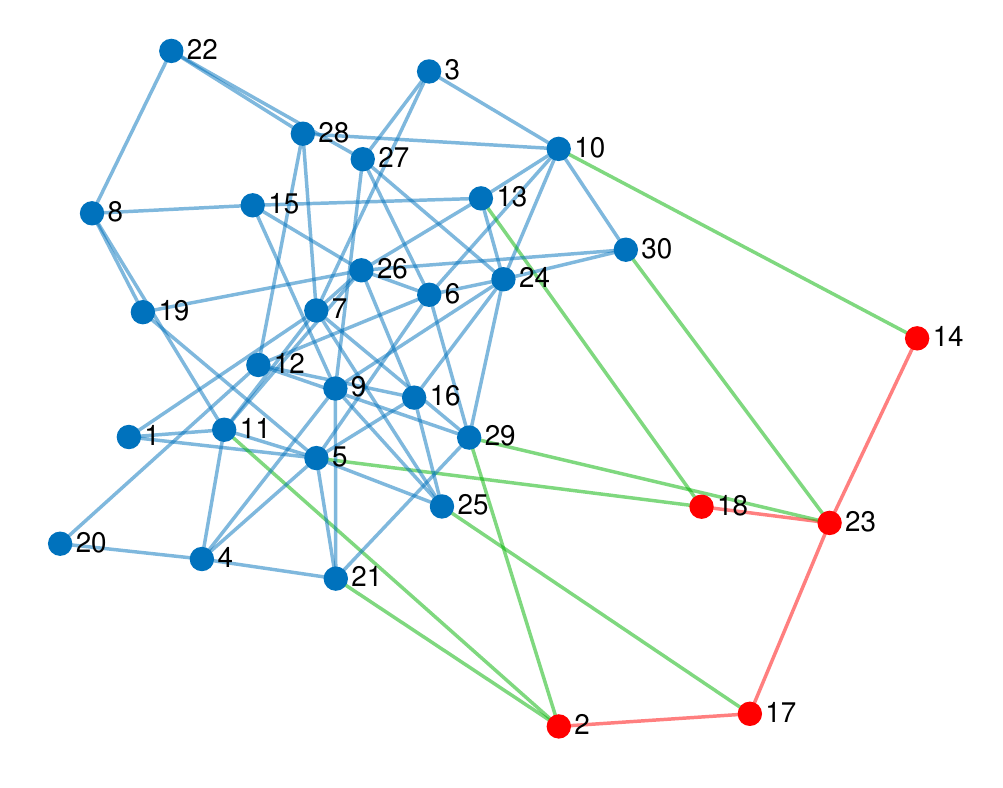}
        \label{fig:L_d_random_N_30_reduced_a}}
    \subfloat[]{\includegraphics[max width=0.49\columnwidth]{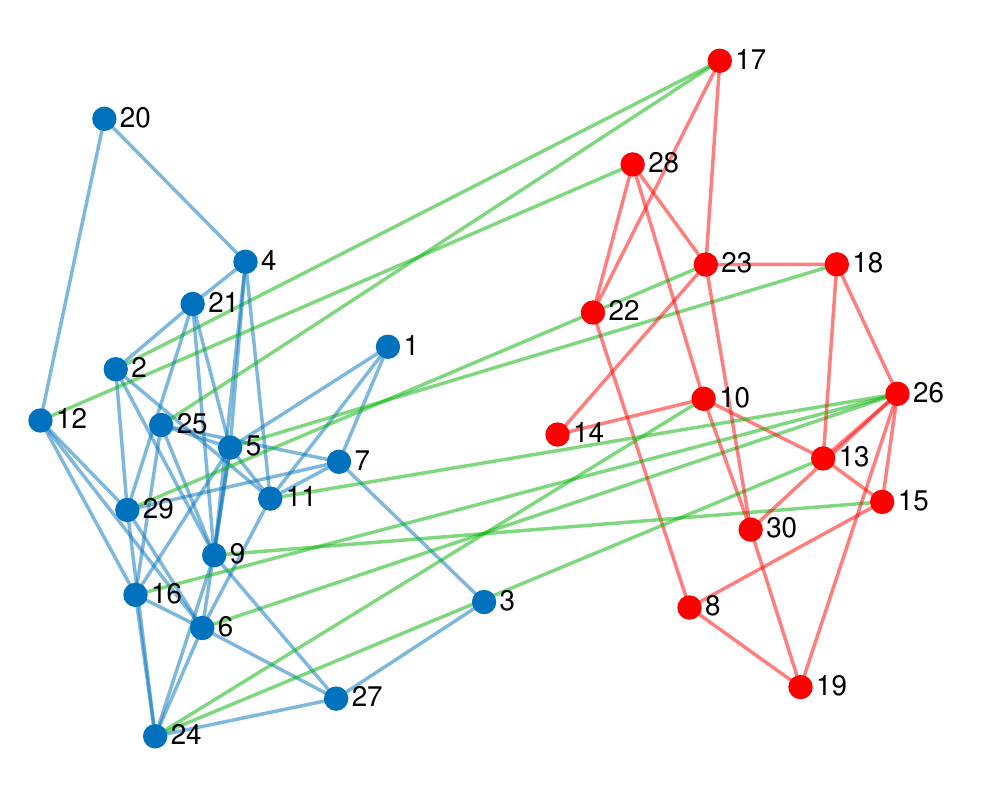}
        \label{fig:L_d_random_N_30_reduced_b}}
    \caption{Graphs {\color{black}$\C{G}_\R{A}$ and $\C{G}_\R{B}$} obtained removing edges from the graph in Figure \ref{fig:L_d_random_N_30_b}. (a) 4 blue edges (2-9, 5-9, 6-11, 6-16) and 4 red edges (17-22, 18-26, 19-30, 23-28) from Figure \ref{fig:L_d_random_N_30_b} were removed; (b) 8 green edges (3-10, 5-19, 6-10, 7-26, 7-28, 8-11, 22-27, 24-30) from Figure \ref{fig:L_d_random_N_30_b} were removed.}
    \label{fig:L_d_random_N_30_reduced}
\end{figure}

Next, we show how the findings in Theorem \ref{thm:discontinuous_coupling} can be used to evaluate the resilience of the network with respect to structural changes in the communication layer.
{\color{black}To this aim, consider the graph in Figure \ref{fig:L_d_random_N_30_a} having 82 edges and its sparsest cut in in Figure \ref{fig:L_d_random_N_30_b}.
    Assume that, due to some fault, 8 edges (roughly 10\% of the total) are removed. 
    We investigate two possible scenarios.
    In scenario A, 4 blue and 4 red \emph{intra-cluster} edges are removed; the new graph in Figure \ref{fig:L_d_random_N_30_reduced_a} has minimum density $\delta_{\C{G}_\R{A}} = 1.080$.
    In scenario B, 8 green \emph{inter-cluster} edges are removed, with the new graph
    in Figure \ref{fig:L_d_random_N_30_reduced_b} having minimum density $\delta_{\C{G}_\R{B}} = 0.747$.
    Consequently, from \eqref{eq:thresholds}, the threshold value $c_{\R{d}}^*$ associated to the latter graph will be larger (thus worse) than that associated to $\C{G}_\R{A}$.
    This suggests that the loss of resilience is greater when the inter-cluster edges (with respect to the sparsest cut of the original graph) are removed, although a detailed analysis of this effect is beyond the scope of this paper and will be the subject of future work.}
%shows that removing some edges rather than others can be more impactful on the synchronizability of the network and hence on its resilience to intentional or unintentional perturbations.
%Specifically, we observe a greater loss of resilience when the inter-cluster edges are removed, although a detailed analysis of this effect is beyond the scope of this paper and will be the subject of future work.
%On the one hand, looking at \eqref{eq:minimum_density}, one could argue that, since $b$ is equal to the number of green edges in \ref{fig:L_d_random_N_30_b}, and the red and blue edges do not contribute to the value of the minimum density, then it is likely that $\delta_{\C{G}_\R{d}, 1} \ge \delta_{\C{G}_\R{d}, 2}$.
%On the other hand, it is still possible that, removing blue and red edges from the graph in Figure \ref{fig:L_d_random_N_30_b}, the sparsest cut changes to a degree that a new particularly critical bottleneck is created in the network, causing a very small minimum density.

%Nonetheless, this intuition, while useful, cannot be generalised a priori to all topologies.}

\section{Proofs}
\label{sec:proofs}

We give here the proofs of the convergence results presented in Section \ref{sec:convergence_results}. 
We start by giving some lemmas and definitions in Section \ref{susec:preliminary_lemmas_and_definitions}, we then introduce the concept of star functions in Section \ref{subsec:star_functions} and show how such functions can be associated to a given graph in Section \ref{subsec:star_function_associated_to_a_graph}. 
In particular, we show that semi-negativity of the star function for a given graph can be studied by assessing its value on the bipartitions of the graph.
This allows to prove Theorem \ref{thm:discontinuous_coupling} in Section \ref{sec:convergence_results}.

\subsection{Preliminary lemmas and definitions}
\label{susec:preliminary_lemmas_and_definitions}

\begin{defn}[Clusterization]\label{def:clusterization}
Given a vector $\B{\xi} \in \BB{R}^n$, we define its \emph{clusterization}, denoted by $\R{clus}(\B{\xi})$, as a partition of the set of indices $\C{I}=\{1, \dots, n\}$, say $\{ \C{I}_{1}, \dots, \C{I}_{Q} \}$ with $1 \le Q \le n$, such that, for all $i, j\in\C{I}$, $\xi_i = \xi_j$ if and only if there exists {\color{black}$q \in (1, \dots, Q)$} such that $i, j \in \C{I}_q$.
\end{defn}

For example, according to this definition, the vector $\B{\xi}=[1\ 1\ 6\ 2\ 2]\T$ has a clusterization $\R{clus}(\B{\xi})=\{ \C{I}_{1}, \C{I}_{2}, \C{I}_{3} \}$ with $\C{I}_{1}=\{1,2\}, \C{I}_{2}=\{3\}, \C{I}_{3}=\{4,5\}$.
Clearly, the clusterization of a vector is unique up to a reordering of the clusters.

\begin{lem}\label{lem:upper_bound_norm_infinite}
Given $\B{\xi} \in \BB{R}^n$ and $\B{A} \in \BB{R}^{n \times n}$, it holds that
$
\B{\xi} \T \B{A} \R{sign}(\B{\xi}) \le {\color{black}\mu_\infty(\B{A})} \left\lVert \B{\xi} \right\rVert_1
$.
\end{lem}
\begin{pf}
Defining $F \triangleq \B{\xi} \T \B{A} \R{sign}(\B{\xi})$, we can write
$
F =\sum_{i=1}^n \left( A_{ii} \left\lvert \xi_i \right\rvert + \sum_{j=1, j\ne i}^n A_{ij} \xi_i \R{sign}(\xi_j) \right)
$,
then 
$
F \le \sum_{i=1}^n \left( {\color{black} A_{ii}} \left\lvert \xi_i \right\rvert + \sum_{j=1, j\ne i}^n \left\lvert A_{ij} \right\rvert \left\lvert \xi_i \right\rvert \right)
$,
and finally 
$
F \le \max_{i = 1, \dots, n} \left( {\color{black} A_{ii} + \sum_{j=1, j \ne i}^{n}\left\lvert A_{ij} \right\rvert} \right)
\sum_{i=1}^n \left\lvert \xi_i \right\rvert.
$
\hspace*{\fill} \qed
%\begin{multline*}
%F =\sum_{i=1}^n \left( A_{ii} \left\lvert \xi_i \right\rvert + \sum_{j=1, j\ne i}^n A_{ij} \xi_i \R{sign}(\xi_j) \right),\\
%\le \sum_{i=1}^n \left( \left\lvert A_{ii} \right\rvert \left\lvert \xi_i \right\rvert + \sum_{j=1, j\ne i}^n \left\lvert A_{ij} \right\rvert \left\lvert \xi_i \right\rvert \right),\\
%\le \max_{i = 1, \dots, n} \left( \sum_{j=1}^{n}\left\lvert A_{ij} \right\rvert \right)
%\sum_{i=1}^n \left\lvert \xi_i \right\rvert.
%\hspace*{\fill} \qed
%\end{multline*}
\end{pf}

\begin{lem}\label{lem:lower_bound_norm_infinite}
Given $\B{\xi} \in \BB{R}^n$ and $\B{A} \in \BB{R}^{n \times n}$, it holds that
$
\B{\xi} \T \B{A} \R{sign}(\B{\xi}) \ge \mu_\infty^-(\B{A}) \left\lVert \B{\xi} \right\rVert_1
$,
with $\mu_\infty^-$ given in {\color{black}Section \ref{sec:notation_and_mathematical_preliminaries}}.
\end{lem}
\begin{pf} Letting again $F \triangleq \B{\xi} \T \B{A} \R{sign}(\B{\xi})$, we have
$
F \ge \sum_{i=1}^n \left( A_{ii}  - \sum_{j=1, j\ne i}^n \left\lvert A_{ij} \right\rvert \right) \left\lvert \xi_i \right\rvert
$,
and 
$
F \ge \min_{i = 1, \dots, n} \left( A_{ii} - \sum_{j = 1, j \ne i}^n \left\lvert A_{ij} \right\rvert \right) \sum_{i=1}^n \left\lvert \xi_i \right\rvert.
$
\hspace*{\fill} \qed

%\begin{multline*}
%F \ge \sum_{i=1}^n \left( A_{ii}  - \sum_{j=1, j\ne i}^n \left\lvert A_{ij} \right\rvert \right) \left\lvert \xi_i \right\rvert,\\   
%\ge \min_{i = 1, \dots, n} \left( A_{ii} - \sum_{j = 1, j \ne i}^n \left\lvert A_{ij} \right\rvert \right) \sum_{i=1}^n \left\lvert \xi_i \right\rvert.
%\hspace*{\fill} \qed
%\end{multline*}
\end{pf}

Let $\BB{V}_d$ be a vector space in $\BB{R}^d$.

\begin{defn}[Cones]\label{def:cones}
\begin{enumerate}[(i)]
\item A set $\C{K} \subseteq \BB{V}_d$ is a \emph{(convex) cone} if, for any $\B{\xi}_1, \B{\xi}_2 \in \C{K}$ and $\alpha_1, \alpha_2 \ge 0$, it holds that $\alpha_1 \B{\xi}_1 + \alpha_2 \B{\xi}_2 \in \C{K}$.
\item A cone is \emph{finitely generated} if it is the {\color{black}\emph{conical combination}} of a finite number of unit norm vectors, which we call \emph{generators} of the cone.
\item A cone $\C{K}$ is \emph{polyhedral} if there exists a matrix $\B{C} = [\B{c}_1 \ \B{c}_2 \ \cdots \ \B{c}_q] \in \BB{R}^{d \times q}$ (with $q \ge d$ and $\B{C}$ having rank $d$) such that $\B{C}\T \B{\xi} \ge \B{0}$, for all $\B{\xi} \in \C{K}$.
\end{enumerate}
\end{defn}

A finitely generated cone in $\BB{R}^3$ is illustrated in Figure \ref{fig:star_function_space_a}. Note that a convex cone contains its boundary.

\begin{lem}[Equivalence of cones \cite{bruns2009polytopes}]\label{lem:cones_equivalence}
A polyhedral cone is a finitely generated cone having $p$ generators; the $i$-th generator $\hat{\B{\xi}}_i$ is such that $\B{c}_j\T \hat{\B{\xi}}_i = 0$ for $n-1$ indices $j \ne i$.
A finitely generated cone is also a polyhedral cone. 
\end{lem}

%---------------------------------------------------------------------
%\subsection{Preliminaries concerning graphs}
%\label{subsec:preliminaries_on_graphs}

\begin{defn}[Incidence matrix \cite{godsil2013algebraic}]
{\color{black}The \emph{incidence matrix} $\B{B}\in \BB{Z}^{N \times N_{\C{E}}}$ of a graph} has columns $\B{b}_i$, $i = 1, \dots, N_{\C{E}}$, where $\B{b}_i$ is associated to edge $i$ connecting vertices $v_j$ and $v_k$, and has all its elements equal to zero, except for positions $j$ and $k$, where it has arbitrarily either $1$ and $-1$ or $-1$ and $1$.
\end{defn}

%{\color{black}
%    \begin{rem}\label{rem:nodes_set}
%        It is possible to define a simple invertible function $\varphi_{\R{v}} : \C{V} \rightarrow \{1, \dots, N \}$ that associates to each node $v_i$ its index $i$.
%        To simplify notation though, we do not discriminate between a vertices set $\C{V}$ and the set $\varphi_{\R{v}}(\C{V})$ of associated indices, and refer to them as if they were the same set.
%\end{rem}}

\begin{defn}[Bipartitions and tripartitions]
{\color{black}We term as \emph{bipartition} (resp. \emph{tripartition}) of a graph $\C{G}$ the partition of the vertices set $\C{V}$ in two (resp. three) subsets, also known as \emph{clusters}, provided that at least two of the clusters are made of connected vertices.}
%We term as \emph{bipartitions} $\hat{\C{B}}$ and \emph{tripartitions} $\hat{\C{T}}$ of a graph $\C{G}$ the sets of all the possible partitions of the vertices set $\C{V}$ of the graph in two or three subsets (or clusters), respectively. 
%We require that in both the bipartitions and the tripartitions there are at least two clusters made of connected vertices.
\end{defn}

We denote generic bipartitions by $\C{B} = \{ \C{I}_1, \C{I}_2 \}$ and tripartitions by $\C{T} = \{ \C{I}_1, \C{I}_2, \C{I}_3\}$, where $\C{I}_i$ is the set of indices of the vertices belonging to the $i$-th cluster.
{\color{black}$\hat{\C{B}}$ and $\hat{\C{T}}$ denote the sets of all possible bipartitions and tripartitions, respectively.}
Finally, {\color{black}we define} $\hat{\C{P}} \triangleq \hat{\C{B}} \cup \hat{\C{T}}$ and a partition that is either a bipartition or a tripartition is denoted by $\C{P}\in \hat{\C{P}}$.

%---------------------------------------------------------------------
\subsection{Star functions}
\label{subsec:star_functions}

\begin{defn}[Star function]\label{def:star_function}
A continuous piecewise-linear function $\phi : \BB{R}^n \rightarrow \BB{R}$ is a \emph{star function} if  (i) it is linear in a set of polyhedral convex cones $\C{K}_j$, $j = 1, \dots, J$, with $J \in \BB{N}_{>0}$, (ii)
the cones can overlap only on their boundaries, and (iii) they are a cover for $\BB{R}^n$.
\end{defn}

\begin{figure}[t]
    \qquad
    \subfloat[][]{\includegraphics[max width=\columnwidth]{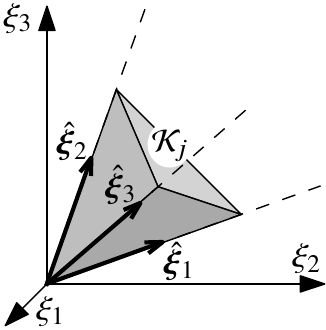}\\
    \label{fig:star_function_space_a}}
    \quad \ \
    \subfloat[][]{\includegraphics[max width=\columnwidth]{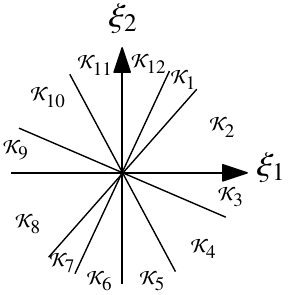}
    \label{fig:star_function_space_b}}
    \caption{(a) A finitely generated cone $\C{K}_j$ in $\BB{R}^3$; $\hat{\B{\xi}}_1$, $\hat{\B{\xi}}_2$, and $\hat{\B{\xi}}_3$ are the generators of the cone. (b) Example of the domain of a star function with 12 cones $\C{K}_j$, $j = 1, \dots, 12$, in the case that $n=2$.}
    \label{fig:star_function_space}
\end{figure}

An example of the domain of a star function is illustrated in Figure \ref{fig:star_function_space_b}.
We can now give the following Lemma, used to assess the semi-negativity of a star function.

\begin{lem}\label{lem:edges}    
Given a star function $\phi : \BB{R}^n \rightarrow \BB{R}$, if $\phi(\B{\xi}) \le 0$ on the generators of the cones $\C{K}_j$, $j = 1, \dots, J$ over which it is defined, then $\phi(\B{\xi}) \le 0$ for all $\B{\xi} \in \BB{R}^n$.
\end{lem}
\begin{pf}
Without loss of generality, consider any cone $\C{K}_j$ where $\phi$ is linear.
Since, by definition $\C{K}_j$ is a finitely generated cone, each of its points can be expressed as 
$
\B{\xi} = \alpha_1 \hat{\B{\xi}}_1 + \alpha_2 \hat{\B{\xi}}_2 + \ldots + \alpha_p \hat{\B{\xi}}_p,% \quad \B{\xi} \in \C{K}_j,
$
where $\hat{\B{\xi}}_1, \dots, \hat{\B{\xi}}_p$ are the $p$ generators of $\C{K}_j$, and $\alpha_1, \dots, \alpha_p \ge 0$.
Then, exploiting linearity of $\phi$, we have
$
\phi(\B{\xi}) = \alpha_1 \phi(\hat{\B{\xi}}_1) + \alpha_2 \phi(\hat{\B{\xi}}_2) + \ldots + \alpha_n \phi(\hat{\B{\xi}}_n)
$,
for all $\B{\xi} \in \C{K}_j$.
Thus, since $\alpha_1, \dots, \alpha_p \ge 0$, if $\phi$ is non-positive on all the generators $\hat{\B{\xi}}_i$ of $\C{K}_j$, then $\phi(\B{\xi}) \le 0$ for all $\B{\xi} \in \C{K}_j$.
%\setminus \{ \B{0} \}$ (the origin is excluded because there $\alpha_1 =  \ldots = \alpha_p = 0$ and $\phi(\B{0}) = 0$).
The same is true for any other $\C{K}_j$.
\hspace*{\fill} \qed
\end{pf}

%{\bf La dimostrazione sopra vale anche per i punti di confine nei quali la funzione star non e' necessariamente lineare secondo la sua definizione?}
%{\color{red} Marco: in effetti $\C{K}_j$ contiene la propria frontiera, quindi sappiamo che la star function è lineare lì. Ho aggiunto una postilla prima del Lemma 11 per chiarire la cosa.}

%---------------------------------------------------------------------
\subsection{Star function associated to a graph}
\label{subsec:star_function_associated_to_a_graph}

Next we give a set of results concerning a specific type of star function that can be associated to a graph $\C{G}$. 
We also show that the properties of this function can be interpreted in a graph-theoretical manner and derive some results that will be useful later in Section \ref{subsec:proof_of_theorem_5} to prove Theorem \ref{thm:discontinuous_coupling}.
We denote by $\C{S}\subset \BB{R}^N$ the subspace
$\label{eq:S}
\C{S} \triangleq \{ \B{e} \in \BB{R}^N  \mid \sum_{i=1}^N e_i = 0 \}
$.
Then, we associate to a graph $\C{G}$ the function $\phi_\C{G} : \C{S} \rightarrow \BB{R}$, given by \begin{equation}\label{eq:phi_G}
\phi_\C{G}(\B{e}) = a_1 \sum_{i = 1}^N \left\lvert \B{i}_i\T \B{e} \right\rvert 
- a_2 \sum_{i = 1}^{N_\C{E}} \left\lvert \B{b}_i\T \B{e} \right\rvert,
\end{equation}
where 
$N$ and $N_\C{E}$ are the numbers of vertices and edges in $\C{G}$, respectively,
$a_1, a_2$ are positive scalars,
$\B{i}_i$ is the $i$-th vector of the canonical basis of $\BB{R}^N$,
and $\B{b}_i$ are the columns of the incidence matrix $\B{B}$ of $\C{G}$.
{\color{black}It is not difficult to show that $\phi_\C{G}$ is indeed a star function, although we omit these derivations for the sake of brevity. 
Each of its cones, say $\C{K}_j$, is the locus where a vector constraint $\B{C}_j\T \B{e} \ge \B{0}$ holds, with 
%$\B{C}_j = [\pm \B{i}_1 \ \cdots \ \pm \B{i}_{N-1} \ \pm \B{i}_{N} \ \pm \B{b}_1 \ \cdots \ \pm \B{b}_{N_{\C{E}}}]$
\begin{equation}\label{eq:constraints_phi_g}
\B{C}_j = \begin{bmatrix}
\pm \B{i}_1 & \cdots & \pm \B{i}_{N} & \pm \B{b}_1 & \cdots & \pm \B{b}_{N_{\C{E}}}
\end{bmatrix},
\end{equation}
and each cone $\C{K}_j$ having a certain combination of plus and minus signs in place of the symbols $\pm$.
We denote by $\hat{\C{H}} \subset \C{S}$ the set of all generators of $\phi_\C{G}$.
}

\begin{lem}\label{lem:partitions}
{\color{black}Let $\C{G}$ be a connected graph, $\phi_\C{G}$ its associated star function defined as in (\ref{eq:phi_G}), and $\hat{\B{e}} \in \hat{\C{H}}$ a generator of $\phi_\C{G}$. 
The clusters $\C{I}_i$ of indices in $\R{clus}(\hat{\B{e}})$ form either a bipartition or a tripartition of $\C{G}$.}%
\footnote{Note that $\R{clus}(\hat{\B{e}})$ is a partition of $\{1, \dots, N \}$.}
\end{lem}
\begin{pf}
To prove the thesis, we need to show that 
%(i) $\R{clus}(\hat{\B{e}}) = \{ \C{I}_1, \dots, \C{I}_Q \}$ is also a partition of $\C{G}$, 
(i) $\R{clus}(\hat{\B{e}}) = \{ \C{I}_1, \dots, \C{I}_Q \}$ with $Q=2$ or $Q=3$; (ii) the partition of $\C{G}$ contains at least two clusters of connected vertices.

%(i) \quad 
%Since $\R{clus}(\hat{\B{e}})$ is a partition of $\{1, \dots, N \}$ (Definition \ref{def:clusterization}), then it is also a partition of the vertices set $\C{V}$ (Remark \ref{rem:vertices_set}) of the graph $\C{G}$, and equivalently a partition of $\C{G}$. 
(i) \quad
From {\color{black}Lemma \ref{lem:cones_equivalence},} it follows that any generator $\hat{\B{e}}$ of $\phi_\C{G}$ is a vector in $\C{S}$ with unit norm such that $N-2$ independent constraints $\B{c}_{i}\T \hat{\B{e}} = 0$ hold.
{\color{black}Recalling \eqref{eq:constraints_phi_g}, we know that }the vectors $\B{c}_i$'s are picked from the set $\C{C}_{\B{i},\B{b}} \triangleq \{ \B{i}_1, \dots, \B{i}_N, \B{b}_1, \dots, \B{b}_{N_\C{E}} \}$.
%we have shown that all the loci $\C{K}_j$ where each argument of the absolute values in \eqref{eq:phi_G} has a certain sign are polyhedral cones.
%Exploiting Lemma \ref{lem:cones_equivalence}, we recall that polyhedral cones are also finitely generated cones, and any generator $\hat{\B{e}}$ is an $(N-1)$-dimensional vector in $\C{S}$ with unit norm such that $N-2$ independent constraints $\B{c}_{i}\T \hat{\B{e}} = 0$ hold, with $\B{c}_i \in \C{C}_{\B{i},\B{b}} \triangleq \{ \B{i}_1, \dots, \B{i}_N, \B{b}_1, \dots, \B{b}_{N_\C{E}} \}$.
We term as $p$ the number of constraints of the form $\B{i}_i\T \hat{\B{e}} = 0$, so that those of the kind $\B{b}_i\T \hat{\B{e}} = 0$ are $N-2-p$, with $0 \le p \le N-2$.
According to the definition of $\B{b}_i$, each of the constraints $\B{b}_i\T \hat{\B{e}} = 0$ implies that two components of $\hat{\B{e}}$ are equal, i.e. that $\hat{e}_j = \hat{e}_k$, for some pair of indices $(j, k)$.
Therefore, from the $N-2-p$ constraints of the form $\B{b}_i\T \hat{\B{e}} = 0$, we can conclude that $\R{clus}(\B{e})$ contains at most $N - (N-2-p) = p+2$ clusters.
%Since $\hat{\B{e}}$ has $N$ elements and we have $N-2-p$ constraints in the form $\hat{e}_j = \hat{e}_k$, there remain $N - (N-2-p) = p+2$ unconstrained elements in $\hat{\B{e}}$, with the others being equal to any one of the $p+2$ unconstrained ones, i.e.~
%\begin{equation}\label{eq:proof_step_08}
%\begin{dcases}
%\hat{e}_j = \hat{e}_k, & \text{iff} \  j, k \in \C{I}_{1}, \\
%\hat{e}_j = \hat{e}_k, & \text{iff} \  j, k \in \C{I}_{2}, \\
%\dots \\
%\hat{e}_j = \hat{e}_k, & \text{iff} \  j, k \in \C{I}_{p+2}.
%\end{dcases}
%\end{equation}
%where $\{\C{I}_1, \C{I}_2, \dots, \C{I}_{p+2}\}$ are sets of indices, which also represent a partition of $p+2$ connected clusters in $\C{G}$; see Remark ***.
We need now to apply the remaining $p$ constraints of the form $\B{i}_i\T \hat{\B{e}} = 0$; we analyse separately the cases when $p=0$ or $p>0$.

\begin{itemize}
\item If $p=0$, there are no constraints like $\B{i}_i\T \hat{\B{e}} = 0$ to consider; thus, $\R{clus}(\hat{\B{e}})=\{\C{I}_1,\C{I}_2\}$ with 
\begin{equation}\label{eq:bipartition}
\begin{dcases}
\hat{e}_j = \epsilon_1, & j \in \C{I}_1, \\
\hat{e}_j = \epsilon_2, & j \in \C{I}_2,
\end{dcases}
\end{equation} 
for some $\epsilon_1, \epsilon_2 \in \BB{R}$, with $\epsilon_1,\epsilon_2 \ne 0$ and $\epsilon_1 \ne \epsilon_2$.
%Note that here $Q = 2$.%
%\footnote{$\epsilon_1$ and $\epsilon_2$ can be determined by exploiting the fact that we must also have $\sum_{i=1}^N \hat{e}_i = 0$ since $\B{e} \in \C{S}$, that $\hat{\B{e}}$ has unit norm and the information on which cone $\hat{\B{e}}$ is a generator for.
%However, this computation does not affect $Q$.
%}

%and $\{ \C{I}_1, \C{I}_2 \}$ is also a bipartition of $\C{V}$ and hence of $\C{G}$; see Figure \ref{fig:bipartition}.

\item If, on the other hand, $p>0$, then  we need to apply the additional $p$ constraints of the form $\B{i}_i\T \hat{\B{e}} = 0$.
Each of these implies an element of $\hat{\B{e}}$ is null.
For example, if $p=1$, we get that $e_i = 0$ for some $i$. 
Without loss of generality, assume $e_1 = 0$ then one cluster in $\R{clus}(\hat{\B{e}})$ will be characterised by all the null elements in $\hat{\B{e}}$ and there will be $Q=p+2=3$ clusters in total.
Analogously, if $p>1$, the $p$ elements such that $e_i=0$ will form one cluster in $\R{clus}(\B{e})$ so that out of the $p+2$ possible clusters in $\R{clus}(\hat{\B{e}})$ only $Q=p+2-(p-1)=3$ will remain. 
Hence, $\R{clus}(\hat{\B{e}})=\{ \C{I}_1, \C{I}_2, \C{I}_3 \}$ with
\begin{equation}\label{eq:tripartition}
\begin{dcases}
\hat{e}_j = \epsilon_{1}, & j \in \C{I}_{1}, \\
\hat{e}_j = \epsilon_{2}, & j \in \C{I}_{2}, \\
\hat{e}_j = 0, & j \in \C{I}_{3},
\end{dcases}
\end{equation}
%with $\{ \C{I}_1, \C{I}_2, \C{I}_3 \}$ being also a tripartition of $\C{V}$; see the representative illustration in Figure \ref{fig:tripartition}.
for some $\epsilon_1, \epsilon_2 \in \BB{R}$, with $\epsilon_1,\epsilon_2 \ne 0$ and $\epsilon_1 \ne \epsilon_2$.
%Again, $\epsilon_{1}$ and $\epsilon_{2}$ can be determined similarly to the case that $p = 0$, without any effect on the value of $Q$.
\end{itemize}

(ii) \quad
To show that $\R{clus}(\hat{\B{e}}) = \{ \C{I}_1, \dots, \C{I}_Q \}$ contains at least two clusters that are clusters of connected vertices in $\C{G}$, it suffices to notice that in our derivation there were at least two clusters induced by the constraints of the form $\B{b}_i\T \hat{\B{e}} = 0$. 
Since by construction the vectors $\B{b}_i$ represent edges in $\C{G}$, then these clusters must correspond to connected vertices in $\C{G}$.
\hspace*{\fill} \qed
\end{pf}

\begin{figure}[t]
    \centering
    \includegraphics[max width=\columnwidth]{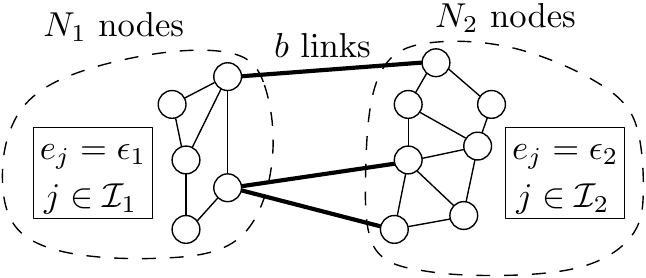}
    \caption{A bipartition $\C{B} = \{ \C{I}_1, \C{I}_2 \}$ of a graph; $N_1$ and $N_2$ are the number of vertices in each cluster and $b$ is the number of edges between the clusters.}
    \label{fig:bipartition}
\end{figure}

\begin{figure}[t]
    \centering
    \includegraphics[max width=\columnwidth]{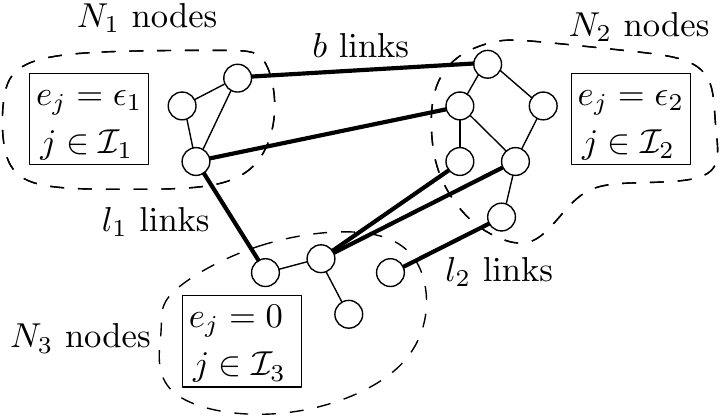}
    \caption{A tripartition $\C{T} = \{ \C{I}_1, \C{I}_2, \C{I}_3 \}$ of a graph; $N_1$, $N_2$, $N_3$ are the number of nodes in each cluster and $b$, $l_1$, $l_2$ are the number of edges between the clusters.}
%        
%        {\color{black}A partition $\C{P}$ of a graph.  If $\C{P}$ is a bipartition, only two clusters exist and $\C{I}_3 = \varnothing$, whereas if $\C{P}$ is a tripartition, $\C{I}_3 \ne \varnothing$. $N_1$, $N_2$, $N_3$ are the number of vertices in each cluster and $b$, $l_1$, $l_2$ are the number of edges between the clusters.}}
    \label{fig:tripartition}
\end{figure}

\begin{figure}[t]
    \centering
    \includegraphics[max width=\columnwidth]{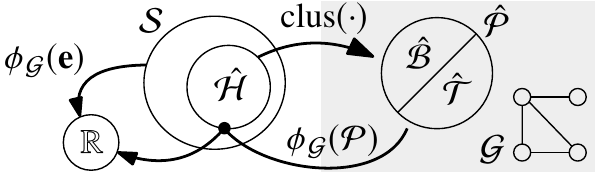}
    \caption{{\color{black}The relations between sets and functions in Section \ref{sec:proofs}.}
%            $\C{G}$ is a graph,
%            $\hat{\C{B}}$ contains all bipartitions of $\C{G}$,
%            $\hat{\C{T}}$ contains all tripartitions of $\C{G}$,
%            $\hat{\C{P}} \triangleq \hat{\C{B}} \cup \hat{\C{T}}$ and
%            $\C{P} \in \hat{\C{P}}$.
%            $\C{S}$ defined in \eqref{eq:S},
%            $\phi_\C{G}$ is a star function \ref{def:star_function} associated to $\C{G}$ and given in \ref{eq:phi_G}, 
%            $\hat{\C{H}}$ contains all generators of $\phi_\C{G}$,
%            $\R{clus}$ denotes the clusterization as given in Definition \ref{def:clusterization}.
%            $\hat{\C{H}}$ is the set of all generators $\hat{\B{e}}$ of $\phi_\C{G}$.
}
    \label{fig:partitions_generators_sets}
\end{figure}

%Next, we show that, under certain conditions, in order to asses the negativity of functions in the form \eqref{eq:star_c}, its sign over just a subset of the generators mentioned in Lemma \ref{lem:edges} needs to be checked.
{\color{black}Given a bipartition $\C{B}$ (resp.~tripartition $\C{T}$) of $\C{G}$, we can always find at least a generator $\hat{\B{e}} \in \hat{\C{H}}$ of $\phi_{\C{G}}$ such that the clusters of indices in $\R{clus}(\hat{\B{e}})$ correspond to the clusters of vertices in $\C{B}$ (resp.~$\C{T}$) verifying (\ref{eq:bipartition}) (resp.~\eqref{eq:tripartition}).
In what follows, we will denote by $\phi_{\C{G}}(\C{B})$ the set of values that the function $\phi_{\C{G}}$ takes over all the vectors $\hat{\B{e}} \in \hat{\C{H}}$ whose clusterization corresponds to $\C{B}$
(analogously for $\phi_{\C{G}}(\C{T})$).
In formal terms, we extend $\phi_{\C{G}}$ to \emph{also} have $\hat{\C{P}}$ as a domain, i.e.~$\phi_{\C{G}} : \hat{\C{P}} \rightrightarrows \BB{R}$ (\eqref{eq:phi_G} still holds), with
$\phi_{\C{G}}(\C{P}) = \{ \phi_{\C{G}}(\hat{\B{e}}) \in \BB{R} \mid \hat{\B{e}} \in \hat{\C{H}} : \R{clus}(\hat{\B{e}}) = \C{P} \}$, for all $\C{P}$ in $\hat{\C{P}}$;
see Figure \ref{fig:partitions_generators_sets}.}

%We will say that $\phi_\C{G}(\C{B}) \le 0$ if this is true for all values of $\phi_\C{G}$ in that set (and equivalently for $\phi_\C{G}(\C{T})$); {\color{black}see Figure \ref{fig:partitions_generators_sets}}.

\begin{lem}\label{lem:biparitions}
Given a connected graph $\C{G}$ and its associated star function $\phi_\C{G}$, if $\phi_\C{G} \le 0$ on all of the bipartitions of $\C{G}$, then $\phi_\C{G} \le 0$ on all of the tripartitions of $\C{G}$.
\end{lem}
\begin{pf}
%Let us denote with $\hat{\B{e}}_\R{b}$ and $\hat{\B{e}}_\R{t}$ the generators of $\phi_\C{G}$ associated to bipartitions and tripartitions of $\C{G}$, respectively.
The proof is composed of three steps. 
First, we determine what conditions must hold so that $\phi_\C{G}(\C{B}) \le 0$ for any bipartition $\C{B}$.
Then, we do the same for a generic tripartition $\C{T}$.
Finally, we show that, for each $\C{T}$, there exists a specific $\C{B}'$ such that $\phi_\C{G}(\C{B}') \le 0 \Rightarrow \phi_\C{G}(\C{T}) \le 0$.
Hence, if $\phi_\C{G}(\C{B}) \le 0$ for all $\C{B} \in \hat{\C{B}}$, then also $\phi_\C{G}(\C{T}) \le 0$ for all $\C{T} \in \hat{\C{T}}$, that is the thesis.
%Then, it is straightforward to show that if $\phi_\C{G} \le 0$ on all bipartitions and tripartitions, then $\phi_\C{G} \le 0$ on all of its generators.
%First, we determine which conditions must hold so that $\phi_\C{G}(\hat{\B{e}}_\R{b}) \le 0$ for any $\hat{\B{e}}_\R{b}$.
%Then we do the same for a generic $\hat{\B{e}}_\R{t}$.
%Finally, we show that for each $\hat{\B{e}}_\R{t}$, there exists a specific $\hat{\B{e}}_\R{b}'$ such that $\phi_\C{G}(\hat{\B{e}}_\R{b}') \le 0 \Rightarrow \phi_\C{G}(\hat{\B{e}}_\R{t}) \le 0$.
%Hence, if $\phi_\C{G}(\hat{\B{e}}_\R{b}) \le 0$ for all $\hat{\B{e}}_\R{b}$, then also $\phi_\C{G}(\hat{\B{e}}_\R{t}) \le 0$ for all $\hat{\B{e}}_\R{t}$, that is $\phi_\C{G}(\hat{\B{e}}) \le 0$ for all generators $\hat{\B{e}}$.

(i) \quad 
Let us consider a generic bipartition $\C{B} = \{ \C{I}_1, \C{I}_2 \}$ of $\C{G}$.
%Thus, $\R{vect}(\C{B})$ is a set of vectors $\B{e} \in \C{S}$ such that \eqref{eq:bipartitions_chi} holds.
Then, from \eqref{eq:phi_G} and \eqref{eq:bipartition}, we can write
%$e_j = \epsilon_1$ for $j \in \C{I}_1$ and $e_j = \epsilon_2$ for $j \in \C{I}_2$, with $\epsilon_1, \epsilon_2$ depending on $\B{e}$ and $\epsilon_1 \ne \epsilon_2$.
%From \eqref{eq:phi_G}, recalling the notation expounded in Remark \ref{rem:notation_partitions}, we have
\begin{equation}\label{eq:proof_step_15}
\phi_\C{G}(\C{B}) = a_1 (N_1 \left\lvert \epsilon_1 \right\rvert + N_2 \left\lvert \epsilon_2 \right\rvert ) - a_2 b \left\lvert \epsilon_1 - \epsilon_2 \right\rvert,
\end{equation}
where $N_1 = \left\lvert \C{I}_1 \right\rvert$, $N_2 = \left\lvert \C{I}_2 \right\rvert$, $b$ is the number of edges connecting vertices in $\C{I}_1$ with vertices in $\C{I}_2$ (see Figure \ref{fig:bipartition}), and $\epsilon_1, \epsilon_2$ are (different non-zero) constants depending on the generic vector $\hat{\B{e}} \in \C{S}$ whose clusterization corresponds to $\C{B}$ according to (\ref{eq:bipartition}).
Even though $N_1$, $N_2$, and $b$ depend on the specific $\C{B}$ being considered, we omit this dependence to simplify the notation. 
%are the sizes of the clusters $\C{I}_1$, $\C{I}_2$ in \eqref{eq:bipartition} (see also Figure \ref{fig:bipartition}), and $b$ is the number of edges between them, computed as $b = \sum_{j \in \C{I}_1} \sum_{k \in \C{I}_2} L_{jk}$, with $L_{jk}$ being the elements of the Laplacian matrix of $\C{G}$.
Since $\sum_{i=1}^N \hat e_i = 0$, then $N_1 \epsilon_1 + N_2 \epsilon_2 = 0$, that is $\epsilon_2 = - \frac{N_1}{N_2} \epsilon_1$.
Therefore, we may rewrite
$
\phi_\C{G}(\C{B}) = 2 a_1 N_1 \left\lvert \epsilon_1 \right\rvert - a_2 b \tfrac{N_1 + N_2}{N_2} \left\lvert \epsilon_1 \right\rvert
$.
Hence, $\phi_\C{G}(\C{B}) \le 0$ if and only if
\begin{equation}\label{eq:proof_step_11}
a_2 \ge \frac{2 a_1 N_1 N_2}{(N_1 + N_2) b},
\end{equation}
independently from the value of the constants $\epsilon_1$ and $\epsilon_2$ associated to the specific vector whose clusterization is being considered.

(ii) \quad
Let us now consider a generic tripartition $\C{T} = \{ \C{I}_1, \C{I}_2, \C{I}_3 \}$.
%$\R{vect}(\C{T})$ is a subset of $\C{S}$ such that \eqref{eq:tripartitions_chi} holds: $e_j = \epsilon_1$ for $j \in \C{I}_1$, $e_j = \epsilon_2$ for $j \in \C{I}_2$, and $e_j = 0$ for $j \in \C{I}_3$, with again $\epsilon_1, \epsilon_2$ depending on $\B{e}$ and $\epsilon_1 \ne \epsilon_2$.
Using similar arguments to those presented for bipartitions, from \eqref{eq:phi_G} and \eqref{eq:tripartition}, we obtain
\begin{multline}\label{eq:proof_step_10}
\phi_\C{G}(\C{T}) = a_1 (
N_1 \left\lvert \epsilon_1 \right\rvert + 
N_2 \left\lvert \epsilon_2 \right\rvert +
N_3 \left\lvert 0 \right\rvert ) \\
- a_2 (
b \left\lvert \epsilon_1 - \epsilon_2 \right\rvert +
l_1 \left\lvert \epsilon_1 - 0 \right\rvert +
l_2 \left\lvert \epsilon_2 - 0 \right\rvert ),
\end{multline}
where $N_1 = \left\lvert \C{I}_1 \right\rvert$, $N_2 = \left\lvert \C{I}_2 \right\rvert$, $N_3 = \left\lvert \C{I}_3 \right\rvert$, and $b$, $l_1$, $l_2$ are the numbers of edges connecting vertices in $\C{I}_1$ with vertices in $\C{I}_2$, vertices in $\C{I}_1$ with vertices in $\C{I}_3$, and vertices in $\C{I}_2$ with vertices in $\C{I}_3$, respectively (see Figure \ref{fig:tripartition}).
$N_1$, $N_2$, $N_3$, $b$, $l_1$, and $l_2$ all depend on $\C{T}$.
%\footnote{Recall that all these values depend on the specific generator/partition being considered, thus the values of $N_1$, $N_2$ and $b$ in \eqref{eq:proof_step_10} are different with respect to those in \eqref{eq:proof_step_09}-\eqref{eq:proof_step_11}.}
Since $\sum_{i=1}^N \hat{e}_i = 0$, then $N_1 \epsilon_1 + N_2 \epsilon_2 + N_3 \cdot 0 = 0$, that is again $\epsilon_2 = - \frac{N_1}{N_2} \epsilon_1$. 
In view of this, and multiplying both sides of \eqref{eq:proof_step_10} by $N_2$, we obtain
$
N_2 \phi_\C{G}(\C{T}) =
2 a_1 N_1 N_2 \left\lvert \epsilon_1 \right\rvert
- a_2 (
b (N_1 + N_2) \left\lvert \epsilon_1 \right\rvert +
l_1 N_2 \left\lvert \epsilon_1 \right\rvert +
l_2 N_1 \left\lvert \epsilon_2 \right\rvert )
$.
As $N_2 > 0$, this yields that $\phi_\C{G}(\C{T}) \le 0$ if and only if
\begin{equation}\label{eq:proof_step_12}
a_2 \ge \frac{2 a_1 N_1 N_2}{b(N_1 + N_2) + N_2 l_1 + N_1 l_2}.
\end{equation} 

(iii) \quad
%Starting from the tripartition associated to a generic $\hat{\B{e}}_\R{t}$, we show that there exists a bipartition associated to some generator $\hat{\B{e}}_\R{b}'$ such that $\phi_\C{G}(\hat{\B{e}}_\R{b}') \le 0 \Rightarrow \phi_\C{G}(\hat{\B{e}}_\R{t}) \le 0$.
We now show that
\begin{equation}\label{eq:proof_step_03}
\forall \C{T} \in \hat{\C{T}} \ \exists \C{B}' \in \hat{\C{B}}
\ \text{s.t.} \
\phi_\C{G}(\C{B}') \le 0 \Rightarrow \phi_\C{G}(\C{T}) \le 0.
\end{equation}
Let us consider again the generic tripartition $\C{T} = \{ \C{I}_1, \C{I}_2, \C{I}_3 \}$ introduced in point (ii); see Figure \ref{fig:tripartition} for a graphical representation.
With an appropriate labelling of the clusters, without loss of generality we can assume that $l_2 \ge l_1$.
To any $\C{T}$, we can always associate a specific bipartition $\C{B}'= \{ \C{I}_1', \C{I}_2' \}$, where $\C{I}_1' = \C{I}_1$ and $\C{I}_2' = \C{I}_2 \cup \C{I}_3$, characterised by $N_1' = \left\lvert \C{I}_1' \right\rvert$, $N_2' = \left\lvert \C{I}_2' \right\rvert$, and $b'$ being the number of edges between $\C{I}_1'$ and $\C{I}_2'$. 
Thus, it follows that $N_1' = N_1$, $N_2' = N_2 + N_3$, and $b' = b + l_1$.
According to \eqref{eq:proof_step_11}, $\phi_\C{G}(\C{B}') \le 0$ if and only if
\begin{equation}\label{eq:proof_step_14}
a_2 \ge
\frac{2 a_1 N_1' N_2'}{(N_1' + N_2') b'} =
\frac{2 a_1 N_1 (N_2 + N_3)}{(N_1 + N_2 + N_3) (b + l_1)}.
\end{equation}
Next, we prove that \eqref{eq:proof_step_14} implies \eqref{eq:proof_step_12}, independently of $\C{T}$.
%From the hypothesis, we know \eqref{eq:proof_step_14} to be true for all $\C{B}' \in \hat{\C{B}}$.
%Then, to conclude the we can show that \eqref{eq:proof_step_14} implies \eqref{eq:proof_step_12}, for all $\C{T} \in \hat{\C{T}}$, that is the thesis, i.e.
%\begin{equation}
%\phi_\C{G}(\C{B}) \le 0 \ \forall \C{B} \in \hat{\C{B}}
%\quad \Rightarrow \quad
%\phi_\C{G}(\C{T}) \le 0 \ \forall \C{T}  \in \hat{\C{T}}.
%\end{equation} 
%To verify that \eqref{eq:proof_step_14} implies \eqref{eq:proof_step_12}
To that aim, we need to show that
\begin{equation*}
\frac{2 N_1 (N_2 + N_3)}{(N_1 + N_2 + N_3) (b + l_1)} \ge 
\frac{2 N_1 N_2}{b(N_1 + N_2) + N_2 l_1 + N_1 l_2}, 
\end{equation*}
which is trivially verified by recalling that $l_2 \ge l_1$. The thesis directly follows.
%or equivalently, exploiting the fact that $N_1 + N_2 + N_3 = N$, that $\eta \ge 0$ with
%\begin{equation*}
%\begin{aligned}    
%\eta &\triangleq (N - N_1) (b(N_1 + N_2) + N_2 l_1 + N_1 l_2) \\
%&\phantom{{}=}- N (b + l_1) N_2 \\
%%&= N N_1 b + N N_2 b + N N_2 l_1 + N N_1 l_2 - N_1^2 b \\
%%&\phantom{{}=}- N_1 N_2 b - N_1 N_2 l_1 - N_1^2 l_2 - N N_2 b - N N_2 l_1 \\
%&= N N_1 b + N N_1 l_2 - N_1^2 b - N_1 N_2 b - N_1 N_2 l_1 - N_1^2 l_2 \\
%%&= Nb - N_1 b - N_2 b - N_2 l_1 + N l_2 - N_1 l_2 \\
%&= (N-N_1 - N_2)b - N_2 l_1 + (N - N_1) l_2.
%\end{aligned}
%\end{equation*}
%Now, recalling that $l_2 \ge l_1$, we have
%\begin{equation*}
%\begin{aligned}
%\eta \ge N_3 b + (N - N_1 - N_2) l_1
%= N_3 (b + l_1) 
%\end{aligned}
%\end{equation*}
%which is certainly non-negative and the thesis follows.
%Now, recall that each generator of $\phi_\C{G}$ is associated to either a bipartition or a tripartition of $\C{G}$ (Lemma \ref{lem:partitions}), and evaluating $\phi_\C{G}$ on a generator is equivalent to evaluating it on the associated partition (Remark \ref{rem:equivalence_generators_partitions}). 
%Therefore, since $\phi_\C{G} \le 0$ on all bipartitions and all tripartitions, then $\phi_\C{G} \le 0$ on all the generators.
\hspace*{\fill} \qed
\end{pf}

We are now ready to give the final result that summarises previous findings and will be used in the proof of the main theorem.
\begin{lem}\label{lem:bipartitions_global_negativity}
If $\phi_\C{G} \le 0$ on all the bipartitions of $\C{G}$, then $\phi_\C{G} \le 0$ for all $\B{e} \in \C{S}$.
\end{lem}
\begin{pf}
According to Lemma \ref{lem:biparitions}, since $\phi_\C{G}(\C{B}) \le 0$ for all $\C{B} \in \hat{\C{B}}$, then 
\begin{equation}\label{eq:proof_step_13}
\phi_\C{G}(\C{P}) \le 0, \quad \forall \ \C{P} \in \hat{\C{P}}.
\end{equation}
Exploiting Lemma \ref{lem:partitions}, the clusterization $\R{clus}(\hat{\B{e}})$ of each generator $\hat{\B{e}} \in \hat{\C{H}}$ is a partition $\C{P} \in \hat{\C{P}}$. 
%
%In other terms, for all $\hat{\B{e}} \in \hat{\C{H}}$, there exists a $\C{P} \in \hat{\C{P}}$ such that $\hat{\B{e}} \in \R{vect}(\C{P})$; see Definition \ref{def:chi_function}.
%Consequently $\phi_\C{G}(\hat{\B{e}}) \in \phi_\C{G}(\R{vect}(\C{P})) = \phi_\C{G}(\C{P})$.
Therefore, \eqref{eq:proof_step_13} implies that
\begin{equation}\label{eq:proof_step_16}
\phi_\C{G}(\hat{\B{e}}) \le 0, \quad \forall \ \hat{\B{e}} \in \hat{\C{H}}.
\end{equation}
Recalling that $\phi_\C{G}$ is a star function, \eqref{eq:proof_step_16} implies the thesis through Lemma \ref{lem:edges}.
\hspace*{\fill} \qed
\end{pf}

\subsection{Proof of Theorem \ref{thm:discontinuous_coupling}}
\label{subsec:proof_of_theorem_5}

%The dynamics of the average state $\tilde{\B{x}}$ of network \eqref{eq:network} under the multilayer control action \eqref{eq:discontinuous_coupling_network} are given by
%\begin{equation}\label{eq:proof_step_05}
%\begin{aligned}
%\dot{\tilde{\B{x}}} = &\frac{1}{N} \sum_{i=1}^N \B{f}(\B{x}_i; t) +
%\sum_{i=1}^{N} \sum_{j=1}^{N} L_{ij} \B{\Gamma} (\B{x}_j - \B{x}_i)\\ 
%&+ \sum_{i=1}^{N} \sum_{j=1}^{N} L_{ij}^\R{d} \B{\Gamma}_\R{d} \R{sign} (\B{x}_j - \B{x}_i).
%\end{aligned}
%\end{equation}    
%As $\B{L}$ and $\B{L}_\R{d}$ are symmetric, the last two terms of the right-hand side of \eqref{eq:proof_step_05} are zero, and therefore we have
The dynamics of the average state $\tilde{\B{x}}$ of network \eqref{eq:network} under the multilayer control action \eqref{eq:discontinuous_coupling_network} are given by
 $\dot{\tilde{\B{x}}} = \frac{1}{N} \sum_{i=1}^N \B{f}(\B{x}_i; t)$.
Therefore, the synchronization error $\B{e}_i$ evolves according to
\begin{multline}
\dot{\B{e}}_i = \dot{\B{x}}_i - \dot{\tilde{\B{x}}}
=\B{f}(\B{x}_i; t) - \frac{1}{N}\sum_{i=1}^{N}\B{f}(\B{x}_i; t) \\
-c \sum_{j=1}^N L_{ij} \B{\Gamma} \B{e}_j 
- c_\R{d} \sum_{j = 1}^{N} L_{ij}^\R{d} \B{\Gamma}_\R{d} \R{sign}(\B{e}_j - \B{e}_i),
\end{multline}    
where we used the fact that $\sum_{j=1}^N L_{ij} \left( \B{x}_j - \B{x}_i \right) = \sum_{j=1}^N L_{ij} \B{x}_j = \sum_{j=1}^N L_{ij} \B{e}_j$ and that $\R{sign}(\B{x}_j - \B{x}_i) = \R{sign}(\B{e}_j - \B{e}_i)$.
Now, consider the candidate common Lyapunov function $V \triangleq \frac{1}{2} \sum_{i=1}^N \B{e}_i\T \B{P} \B{e}_i$. 
Its time derivative is $\dot{V} = \sum_{i=1}^N \B{e}_i\T \B{P} \dot{\B{e}}_i$, that is, 
\begin{multline}\label{eq:dotv}
\dot{V} = \sum_{i=1}^N \B{e}_i\T \B{P} \left( \B{f}(\B{x}_i; t) - \frac{1}{N}\sum_{i=1}^{N}\B{f}(\B{x}_i; t) \right)
-c \sum_{i=1}^N \sum_{j=1}^N \\
L_{ij} \B{e}_i\T \B{P} \B{\Gamma} \B{e}_j
- c_\R{d} \sum_{i=1}^N \sum_{j = 1}^{N} L_{ij}^\R{d} \B{e}_i\T \B{P} \B{\Gamma}_\R{d} \R{sign}(\B{e}_j - \B{e}_i).
\end{multline}
As $\sum_{i=1}^{N} \B{e}_i = 0$, we have $\sum_{i=1}^{N} \B{e}_i\T \B{P} \B{f}(\tilde{\B{x}}; t) = 0$ and $\sum_{i=1}^{N} \B{e}_i\T \B{P} \left( \sum_{i=1}^{N} \B{f}(\B{x}_i; t) / N \right) = 0$.
Thus, we can rewrite \eqref{eq:dotv} as
\begin{multline}
\dot{V} = \sum_{i=1}^N \B{e}_i\T \B{P} \left[ \B{f}(\B{x}_i; t) - \B{f}(\tilde{\B{x}}; t) \right]
-c \sum_{i=1}^N \sum_{j=1}^N L_{ij} \B{e}_i\T \B{P} \B{\Gamma} \B{e}_j \\
- c_\R{d} \sum_{(i,j) \in \C{E}_\R{d}}  (\B{e}_i - \B{e}_j)\T \B{P} \B{\Gamma}_\R{d} \R{sign}(\B{e}_i - \B{e}_j),
\end{multline}
where we also exploited that, since $L_{ij}^\R{d} = L_{ji}^\R{d}$, for each term $\B{e}_i\T \B{P} \B{\Gamma}_\R{d} \R{sign}(\B{e}_j - \B{e}_i)$ there exists another term $\B{e}_j\T \B{P} \B{\Gamma}_\R{d} \R{sign}(\B{e}_i - \B{e}_j)$; in addition, we recall that $\C{E}_\R{d}$ is the set of edges in the graph $\C{G}_\R{d}$.
%Hence, we may recast $\dot{V}$ as
%\begin{multline}
%\dot{V} = \sum_{i=1}^N \B{e}_i\T \B{P} \left[ \B{f}(\B{x}_i; t) - \B{f}(\tilde{\B{x}}; t)\right]
%-c \sum_{i=1}^N \sum_{j=1}^N L_{ij} \B{e}_i\T \B{P} \B{\Gamma} \B{e}_j \\
%- c_\R{d} \sum_{(i,j) \in \C{E}_\R{d}}  (\B{e}_i - \B{e}_j)\T \B{P} \B{\Gamma}_\R{d} \R{sign}(\B{e}_i - \B{e}_j),
%\end{multline}
%recalling that $\C{E}_\R{d}$ is the set of edges in the graph $\C{G}_\R{d}$.
Then, we use the hypothesis that $\B{f}$ is \textsigma-QUAD and get
\begin{multline}\label{eq:proof_step_01}
\dot{V} \le  \sum_{i=1}^N \left( \B{e}_i\T \B{Q} \B{e}_i + \B{e}_i\T \B{M} \R{sign}(\B{e}_i) \right)
- \hspace{-0.05cm} c \sum_{i=1}^N \sum_{j=1}^N  L_{ij} \\
\B{e}_i\T \B{P} \B{\Gamma} \B{e}_j - c_\R{d} \sum_{(i,j) \in \C{E}_\R{d}}  (\B{e}_i - \B{e}_j)\T \B{P} \B{\Gamma}_\R{d} \R{sign}(\B{e}_i - \B{e}_j).
\end{multline}
Now, if we define $\bar{\B{y}} \triangleq \left(\B{B}_\R{d}\T \otimes \B{I}_n \right) \bar{\B{e}}$, where $\B{B}_\R{d}$ is the incidence matrix of $\C{G}_\R{d}$, then we can rewrite \eqref{eq:proof_step_01} as
%\begin{multline}
%\dot{V} \le \bar{\B{e}}\T \left( \B{I}_N \otimes \B{Q} - c \B{L} \otimes \B{P} \B{\Gamma} \right) \bar{\B{e}} \\
%+ \bar{\B{e}}\T \left( \B{I}_N \otimes \B{M} \right) \R{sign}(\bar{\B{e}}) 
%- c_\R{d} \bar{\B{y}} \left( \B{I}_N \otimes \B{P} \B{\Gamma}_\R{d} \right) \R{sign} (\bar{\B{y}}).
%\end{multline}
%Then, we recast 
$\dot{V} \le W_1 + W_2$, where 
\begin{equation}\label{eq:W_1}
W_1 \triangleq \bar{\B{e}}\T \left( \B{I}_N \otimes \B{Q} - c \B{L} \otimes \B{P} \B{\Gamma} \right) \bar{\B{e}},
\end{equation}
\begin{equation}\label{eq:W_2}
W_2 \triangleq \bar{\B{e}}\T \left( \B{I}_N \otimes \B{M} \right) \R{sign}(\bar{\B{e}}) 
- c_\R{d} \bar{\B{y}}{\color{black}\T} \left( \B{I}_{\color{black}N_{\C{E}_\R{d}}} \otimes \B{P} \B{\Gamma}_\R{d} \right) \R{sign} (\bar{\B{y}}).
\end{equation}

We can then study $W_1$ and $W_2$ separately, so as to find conditions that guarantee the former is negative definite and the latter is semi-negative definite.

%---------------------------------------------------------------------
\subsubsection{Negativity of $W_1$}

To find a condition such that $W_1 < 0$, {\color{black}note that $\lambda_1(\B{L}) = 0$ with corresponding eigenvector $\B{1}_N$ \cite[§ 13.1]{godsil2013algebraic}, and let $\B{G} \triangleq \frac{\B{P} \B{\Gamma} + (\B{P} \B{\Gamma})\T}{2}$.
Then, by \cite[Thm. 4.2.12]{horn1991topics}, the first $n$ smallest eigenvalues of $\B{L} \otimes \B{G}$ are all 0, and the $n$  corresponding eigenvectors are $\B{1}_N \otimes \B{w}_i$, where $\B{w}_i$, $i = 1, \dots, n$ are the eigenvectors of $\B{G}$.
Notice that by construction $\bar{\B{e}}$ is orthogonal to all these eigenvectors, as $\bar{\B{e}}\T (\B{1}_N \otimes \B{v}) = 0$ for any $\B{v} \in \BB{R}^n$.
Therefore, through \cite[Thm. 4.2.2]{horn2012matrix} we can write
\begin{equation*}
	\lambda_{n+1} \left( \B{L} \otimes \B{G} \right) = 
	\min_{\bar{\B{e}} : \, \bar{\B{e}}\T (\B{1}_N \otimes \B{v}) = 0, \, \bar{\B{e}} \ne 0}
	\frac{\bar{\B{e}}\T \left( \B{L} \otimes \B{P} \B{\Gamma} \right) \bar{\B{e}}}{ \bar{\B{e}}\T \bar{\B{e}} }.
\end{equation*}
Hence, as $\lambda_2(\B{L})\lambda_1 \left( \B{G} \right) = \lambda_{n+1} \left( \B{L} \otimes \B{G} \right)$, we get that 
$- c \bar{\B{e}}\T \left( \B{L} \otimes \B{P} \B{\Gamma} \right) \bar{\B{e}} \le - c \lambda_2(\B{L}) \mu_2^-(\B{P} \B{\Gamma}) \left\lVert \bar{\B{e}} \right\rVert_2^2$.
Moreover, it holds that 
$\bar{\B{e}}\T \left( \B{I}_N \otimes \B{Q} \right) \bar{\B{e}} \le \left\lVert \bar{\B{e}} \right\rVert_2^2 \lambda_{\R{max}}(\frac{
    \B{Q} + \B{Q}\T}{2})$.
Thus, in \eqref{eq:W_1},} $W_1 < 0$ if $c > c^*$, with $c^*$ defined as in \eqref{eq:thresholds}.
Note that the fact that $\C{G}$ is connected ensures that $\lambda_{2} (\B{L}) > 0$.

%---------------------------------------------------------------------
\subsubsection{Semi-negativity of $W_2$}
\label{sec:semi_negativity_of_W_2}

Next, we seek an expression of the threshold value $c_\R{d}^*$ such that $W_2 \le 0$ if $c_\R{d} \ge c_\R{d}^*$.
Firstly, consider that, from \eqref{eq:W_2}, using Lemmas \ref{lem:upper_bound_norm_infinite} and \ref{lem:lower_bound_norm_infinite}, we have
\begin{equation}\label{eq:proof_step_02}
W_2 \le \left\lVert \bar{\B{e}} \right\rVert_1 {\color{black}\mu_\infty(\B{M})}
- c_\R{d} \left\lVert \bar{\B{y}} \right\rVert_1 \mu_\infty^-( \B{P} \B{\Gamma}_\R{d}).
\end{equation}
Using the definition of the vector 1-norm, we have 
$\lVert \bar{\B{e}} \rVert_1 
= \sum_{i = 1}^{nN} \lvert \bar{e}_i \rvert 
= \sum_{h = 1}^{n} \lVert \B{e}^h \rVert_1
= \sum_{h = 1}^{n} \sum_{i = 1}^{N} \lvert \B{i}_i\T \B{e}^h \rvert$,
where $\B{i}_i$ and $\B{e}^h$ are defined in Section \ref{sec:notation_and_mathematical_preliminaries}.
Note that $\B{e}^h \in \C{S}$, with $\C{S}$ being defined in Section \ref{subsec:star_function_associated_to_a_graph}.
Similarly, it is straightforward to compute that
$\left\lVert \bar{\B{y}} \right\rVert_1 
= \sum_{h = 1}^{n} \sum_{i = 1}^{N_{\C{E}_\R{d}}} \left\lvert \B{b}_i\T \B{e}^h \right\rvert$,
where $\B{b}_i$ are the columns of the incidence matrix $\B{B}_\R{d}$ of $\C{G}_\R{d}$.
For  the sake of compactness, we define $M \triangleq {\color{black}\mu_\infty(\B{M})}$ and ${\color{black}\gamma} \triangleq \mu_\infty^-( \B{P} \B{\Gamma}_\R{d})$.
Thus, we can recast \eqref{eq:proof_step_02} as $W_2 \le \sum_{h=1}^{n} W_2^h$, where 
\begin{equation}\label{eq:W2h}
W_2^h (\B{e}^h)\triangleq 
M \sum_{i = 1}^{N} \left\lvert \B{i}_i\T \B{e}^h \right\rvert
- c_{\R{d}} {\color{black}\gamma} \sum_{i = 1}^{N_{\C{E}_\R{d}}} \left\lvert \B{b}_i\T \B{e}^h \right\rvert.
\end{equation}

The analytical framework and results presented in Sections \ref{subsec:star_functions} and \ref{subsec:star_function_associated_to_a_graph} can be used to more easily assess the semi-negativity of $W_2^h$.
In fact, $W_2^h$ is in the form \eqref{eq:phi_G}, and thus is a star function associated to the graph $\C{G}_\R{d}$; see Definition \ref{def:star_function}. %and Lemma \ref{lem:phi_G}.
Exploiting Lemma \ref{lem:bipartitions_global_negativity}, it is immediate to state that $W_2^h(\B{e}^h) \le 0$ for all $\B{e}^h \in \C{S}$ {\color{black}(that is, globally)} if $W_2^h(\C{B}) \le 0$ for all $\C{B} \in \hat{\C{B}}$; $\hat{\C{B}}$ being the set of all bipartitions of $\C{G}_\R{d}$.

Consider a generic bipartition $\C{B} = \{ \C{I}_1, \C{I}_2 \}$ of $\C{G}_\R{d}$, where $\C{I}_1$ and $\C{I}_2$ are the indices of the vertices in the two connected clusters.
Moreover, let $N_1 = \left\lvert \C{I}_1 \right\rvert$, $N_2 = \left\lvert \C{I}_2 \right\rvert$, and $b$ be the number of edges connecting a vertex in $\C{I}_1$ with a vertex in $\C{I}_2$ (see Figure \ref{fig:bipartition}); note that $N_1$, $N_2$, and $b$ depend on $\C{B}$.
According to \eqref{eq:proof_step_15} and \eqref{eq:proof_step_11}, $W_2^h(\C{B}) \le 0$ if and only if
\begin{equation}\label{eq:proof_step_07}
c_\R{d} \ge \frac{2 M}{N {\color{black}\gamma}} \left( \frac{N_1 N_2}{b} \right),
\end{equation}
where we used the fact that  $N_1 + N_2 = N$.
We highlight that this last step is independent from $h$; therefore, if \eqref{eq:proof_step_07} holds, then $W_2^h(\C{B}) \le 0 \ \forall h = 1, \dots, n$.
From the hypotheses we know that 
\begin{equation}\label{eq:proof_step_04}
c_\R{d} \ge c_\R{d}^* \triangleq \frac{1}{\delta_{\C{G}_\R{d}}}\frac{M}{{\color{black}\gamma}} = 
\frac{2 M}{N {\color{black}\gamma}} \frac{1}{\min_{\C{C} \in \hat{\C{C}}_{\C{G}_\R{d}}} \left( \frac{b}{N_1 N_2} \right)}.
\end{equation}
where $\delta_{\C{G}_\R{d}}$ is the minimum density of $\C{G}_\R{d}$ (Definition \ref{def:minimum_density}) and $\hat{\C{C}}_{\C{G}_\R{d}}$ is the set of all possible cuts on $\C{G}_\R{d}$.
Thus, \eqref{eq:proof_step_04} can be reformulated as 
\begin{equation}
c_\R{d} \ge 
\frac{2 M}{N {\color{black}\gamma}} \frac{1}{\min_{\C{B} \in \hat{\C{B}}} \left( \frac{b}{N_1 N_2} \right)}
= \frac{2 M}{N {\color{black}\gamma}} \max_{\C{B} \in \hat{\C{B}}} \left( \frac{N_1 N_2}{b} \right).
\end{equation}
Therefore, \eqref{eq:proof_step_07} holds for all $\C{B} \in \hat{\C{B}}$.
This ensures that $W_2^h(\C{B}) \le 0$ for all $\C{B} \in \hat{\C{B}}$, which through Lemma \ref{lem:bipartitions_global_negativity} gives $W_2^h \le 0$ globally.
As mentioned previously, if $W_2^h \le 0$ for some $h$, then $W_2^h \le 0$ for all $h$, and hence $W_2 \le 0$.
This completes the proof, as $\dot{V} = W_1 + W_2 < 0$ globally.
\hspace*{\fill} \qed

{\color{black}
We note that the occurrence of sliding dynamics does not impact the analysis, which is conducted using a common Lyapunov function method, with $V$ being a valid Lyapunov function in all the state space of the network, even in the regions where there is sliding (if any).
}

\section{Conclusions}
\label{sec:conclusions}
We addressed the challenging problem of proving global asymptotic convergence to synchronization in  a network of piecewise-smooth dynamical systems, without employing, as done in previous attempts in the literature, costly centralised control actions on all the nodes.
We showed that, under some assumptions on the agents' vector field, adding a discontinuous coupling layer to the commonly used diffusive coupling protocol is sufficient to ensure convergence.

We derived sufficient conditions that allow computation of the critical values of the coupling gains required for convergence. %, even when the inner coupling matrices are not positive definite.
The conditions depend explicitly on structural properties of the underlying network graphs that can be computed algorithmically. 
In particular, we introduced the concept of \emph{minimum density} of a graph that can be used to compute the critical coupling gain of the discontinuous control layer.

An open problem left for further study is to investigate if there exist some best structures of the diffusive and discontinuous coupling layers in terms of performance, robustness and stability. For example, preliminary numerical simulations reported in \cite{coraggio2018synchronization} show that different layers' structures can enhance the regions in the control parameter space where synchronization is attained.

% ACKNOWLEDGEMENTS %%%%%%%%%%%%%%%%%%%%%%%%%%%%%%%%%%%%%%%%%%%%%%%%%%%
\begin{ack}
\color{black}{The authors sincerely thank Josep M. Olm (Universitat Politècnica de Catalunya) for his insightful comments and suggestions during the preparation of this manuscript.}
\end{ack}

% BIBLIOGRAPHY %%%%%%%%%%%%%%%%%%%%%%%%%%%%%%%%%%%%%%%%%%%%%%%%%%%%%%%
\bibliographystyle{plain}
\bibliography{automatica_references}

% APPENDIX %%%%%%%%%%%%%%%%%%%%%%%%%%%%%%%%%%%%%%%%%%%%%%%%%%%%%%%%%%%
% Each appendix must have a short title. Sections and subsections are supported in the appendices. 

\end{document}